\documentclass[twocolumn]{aastex631}

\usepackage{hyperref}
\usepackage{graphicx}	
\usepackage{color}
\usepackage{amsmath}

\definecolor{MyDarkRed}{rgb}{0.71,0.14,0.07}

\newcommand{\PhiN}{\Phi_{\text{N}}}
\newcommand{\PhiM}{\Phi_{\text{M}}}
\newcommand{\NPM}{||{\grad}\PhiM||}
\newcommand{\PDM}{\rho_{\text{ph}}}
\newcommand{\ac}{a_0}
\newcommand{\rhoB}{\rho_\text{B}} 
\newcommand{\rhob}{\rho_\text{b}}
\newcommand{\rhod}{\rho_\text{d}} 
\newcommand{\Md}{M_\text{d}}
\newcommand{\Mb}{M_\text{b}} 
\newcommand{\Rd}{R_\text{d}}
\newcommand{\zd}{z_\text{d}}
\newcommand{\rb}{r_\text{b}}

\newcommand{\grad}{\vec{\nabla}}

\begin{document}

\title{A Novel Test for MOND: Gravitational Lensing by Disc Galaxies}

\author[0009-0002-6751-2695]{Christopher Harvey-Hawes}
\email{christopher.harvey-hawes@pg.canterbury.ac.nz}
\affiliation{School of Physical \& Chemical Sciences, University of Canterbury, Private Bag 4800, Christchurch 8140, New Zealand}

\author[0000-0003-2783-3603]{Marco Galoppo}
\email{marco.galoppo@pg.canterbury.ac.nz}
\affiliation{School of Physical \& Chemical Sciences, University of Canterbury, Private Bag 4800, Christchurch 8140, New Zealand}
 
\begin{abstract}

Disc galaxies represent a promising laboratory for the study of gravitational physics, including alternatives to dark matter, owing to the possibility of coupling rotation curves' dynamical data with strong gravitational lensing observations. In particular, Euclid, DES and LSST are predicted to observe hundreds of thousands of gravitational lenses. Here, we investigate disc galaxy strong gravitational lensing in the MOND framework. We employ the concept of equivalent Newtonian systems within the quasi-linear MOND formulation to make use of the standard lensing formalism. We derive the phantom dark matter distribution predicted for realistic disc galaxy models and study the impact of morphological and mass parameters on the expected lensing. We find purely MONDian effects dominate the lensing and generate non-trivial correlations between the lens parameters and the lensing cross section. Moreover, we show that the standard realisation of MOND predicts a substantial increase in the number count of disc galaxy lenses compared to the dark matter-driven predictions, making it distinguishable from the latter in upcoming surveys. Finally, we argue that disc galaxy gravitational lensing, coupled to additional astronomical observations, can be used to constrain the interpolating function of MOND.

\end{abstract}


\section{Introduction}

The dark matter (DM) hypothesis is an essential element of the standard cosmological model, $\Lambda$ Cold Dark Matter ($\Lambda$CDM) \citep[see e.g.,][]{Planck_2020}. DM was originally introduced to explain the observed rotation curves of disc galaxies \citep{Rubin_1978,Bosma_1978,Sofue_2001}: both the lack of the expected quasi-Keplerian fall-off in the rotational velocity curve at large distances and the rotational velocity amplitudes pointed to an additional but invisible source of gravity, i.e., DM.

DM can successfully explain a plethora of astrophysical and cosmological observations, such as the cosmic microwave background power spectrum \citep{Hinshaw_2009} and gravitational lensing by galaxies and galaxy clusters \citep[see e.g.,][]{Treu_2010,Bartelmann_2010}. In particular, DM halos are believed to dominate strong gravitational lensing (SGL) by galaxies \citep{Schneider_2006}. 

However, the results of the many experiments aimed at the direct detection of DM particles are, to date, inconclusive \citep[see, e.g.,][]{Liu_2017,Yue_2021,Blanco_2024}. Furthermore, in the past two decades, the very nature of DM has become a point of strong debate within the scientific community \citep{Bertone_2018,Profumo_2019}.

Various alternative explanations of astrophysical observations have been proposed in the form of novel theories of gravity: modified gravity \citep[MoG,][]{Moffat_2006,Moffat_2013}, fractional gravity \citep{Giusti_2020,Benetti_2023}, rainbow gravity \citep{Magueijo_gravitys_2004}, emergent gravity \citep{Verlinde_2011,Verlinde_2017}, postquantum classical gravity \citep{Oppenheim_2023,Oppenheim_2024}, $f(R)$ gravities \citep{Sotiriou_2010}, among others, as well as general relativistic models of disc galaxies \citep{Astesiano_towards_2022,Beordo_geometry-driven_2024,Galoppo_2024a,Galoppo_2024b}.

Any viable alternative to DM must explain the phenomena naturally understood within the standard DM paradigm. Furthermore, observational procedures must be designed to differentiate between the many theories now present in the literature, and rule out those which do not match the observations. In this regard, disc galaxies represent an ideal laboratory to design tests to investigate DM, and differentiate it from its alternatives. Indeed, we recall that in recent years the coupling of dynamics and SGL data in massive disc galaxies has already
been used, on the few systems observed, to accurately probe the nature of DM \citep{Trott_2010,Dutton_2011,Suyu_2012}. 

In this work, we investigate inclination effects in disc galaxy SGL in the MOdified Newtonian Dynamics \citep[MOND,][]{Milgrom_modification_1983} scenario (see e.g., \cite{Famaey_2012}, and \cite{Banik_2021} for reviews). To date, MOND represents the most widely studied alternative to the DM paradigm as it has been shown to successfully explain a plethora of astrophysical observations, such as kinematics of galaxies \citep{Famaey_2012,Zhu_2023}, dynamics of wide binary stars \citep{Hernandez_2023, Chae_2024}, radial acceleration relation of galaxies \citep[RAR,][]{Lelli_2017,Mistele_2024b}, orbital velocity of interacting galaxy pairs \citep{Scarpa_2022}, early-universe galaxy and cluster formation \citep{McGaugh_2024}, and galaxy-scale gravitational lensing \citep{Milgrom_2013, Mistele_2024a}.

In the Lagrangian formulation of MOND by \citealp{Bekenstein_1984}, referred to as AQUAL, the linear Poisson equation linking the Newtonian gravitational potential, $\PhiN$, to the matter density, $\rho$, i.e.,
\begin{equation}\label{eq:Poisson}
    \Delta \PhiN  = 4\pi G \rho \, ,
\end{equation}
is replaced by the non-linear partial differential equation (PDE) 
\begin{equation}\label{eq:AQUAL}
    \grad \cdot \left(\mu(x) \grad \PhiM  \right) = 4\pi G \rho \, ,
\end{equation}
where $x = \NPM/\ac$, $\ac \approx 1.2 \times 10^{-13}$ km/s$^2$ is the MOND universal acceleration scale \citep{Famaey_2012}, $\PhiM$ is the Milgromian gravitational potential and $\mu(x)$ is the MOND interpolating function, with the asymptotic limit behaviours
\begin{equation}\label{eq:mu}
    \mu(x) \approx 
    \begin{cases} 
     1, \,\,\, x\gg 1 , \\
     x, \,\,\, x\ll 1 . \\
    \end{cases}
\end{equation}
Eqn.\ \eqref{eq:mu} then implies that for $\NPM \gg \ac$, Eqn.\ \eqref{eq:AQUAL} reduces to \eqref{eq:Poisson}, and MOND recovers the standard Newtonian regime, whilst for $\NPM \ll \ac$ one obtains the deep MOND (dMOND) regime, practically defined by 
\begin{equation}\label{eq:dMOND}
    \grad \cdot\left(\NPM  \, \grad \PhiM\right) = 4\pi G  \rho  \ac \, .
\end{equation}
The extent to which MOND effects dominate an astrophysical system can be roughly gauged by evaluating \citep{Famaey_2012, Banik_2021}
\begin{equation}
    \kappa = \frac{GM}{r_c^2 \ac}\, ,
\end{equation}
where $M$ is the system mass and $r_c$ is the typical scale length of the system. When $\kappa \gg 1$ the system is in a Newtonian regime, whilst for $\kappa \ll 1$ dMOND applies. In particular, we recall that for disc galaxies $\kappa \approx 1$, so that these require the full use of Eqn.\ \eqref{eq:AQUAL} to be described within the MOND scenario.

We note that for every astrophysical system modelled through MOND, it is possible to construct the equivalent Newtonian system (ENS), i.e., a system with the same baryonic matter density, $\rhoB$, supplemented by an additional halo of phantom dark matter (PDM), with density profile $\PDM$, such that $\PhiN = \PhiM$ throughout the physical system. In the AQUAL formulation of MOND, the construction of an ENS is a complex task due to the highly non-linear nature of Eqn.\ \eqref{eq:AQUAL}. 

However, \cite{Milgrom_2010} constructed a different formulation of MOND which allows for a more direct approach to building ENS, i.e., quasi-linear MOND (hereafter QUMOND). QUMOND essentially relies on an approximation of Eqn.\ \eqref{eq:AQUAL} in which $\PhiM$ is replaced by $\PhiN$. By this approach QUMOND involves only the solution of linear PDEs, with only non-linear algebraic steps. Indeed, in QUMOND, the Milgromian gravitational potential now solves for 
\begin{equation}
    \label{eq:QMOND}
    \Delta \PhiM = 4 \pi G \left[\rhoB+\PDM\right]\, ,
\end{equation}
where $\PDM$ is given by
\begin{equation}
    \label{eq:PDM}
    \PDM = \frac{1}{4\pi G}\grad\cdot \left[\tilde{\nu}(y)\grad\PhiN\right]\, ,
\end{equation}
where $y=||\grad \PhiN||/\ac$, and $\tilde{\nu}(y)$ has as asymptotic limiting behaviour 
\begin{equation}\label{eq:nu}
    \tilde{\nu}(y) \approx 
    \begin{cases} 
     0, \,\,\,\,\,\,\,\,\,\,\,\, y\gg 1 , \\
     y^{-1/2}, \,\,\, y\ll 1 . \\
    \end{cases}
\end{equation}
Therefore, within the QUMOND approach, given $\rhoB$, the ENS is straightforwardly built by \textbf{(i)} first solving the classical Poisson equation for the baryonic matter, which becomes the source of $\PDM$ via a non-linear algebraic procedure involving $\tilde{\nu}$; \textbf{(ii)} then by solving once more the Poisson equation for $\rho = \rhoB + \PDM$. 

Here, we stress that the PDM is not constituted by real particles, and thus does not lead to dynamical dissipation or friction \citep{Famaey_2012,Bìlek_2021,Kroupa_2023}. Furthermore, we note that the PDM density can become negative locally and, although it will respect the symmetries of the baryonic distribution, it can present shifts in its concentration peaks w.r.t.\ the former. In particular, for flattened matter distributions such as disc galaxies, the local positivity of $\PDM$ is not assured, nor is the correspondence between the peak of baryonic density and of PDM \citep{Milgrom_1986,Ciotti_2006,Wu_2008,Ciotti_2012,Ko_2016}. Through the use of the ENS, one can study lensing of disc galaxies in MOND, whilst utilising the tools constructed for classical SGL within the $\Lambda$CDM paradigm. This allows us to bypass a series of difficulties in the study of SGL within MOND, e.g., the definition of the thin lens approximation, commonly employed for galaxy-scale SGL \citep{Mortlock_2001}. Furthermore, the ENS allows us to probe the Newtonian and ``transition'' regime of MOND in SGL, whilst naturally accounting for the impossibility of SGL in the dMOND limit \citep[see e.g.,][]{Milgrom_2020}.

In this paper, we investigate inclination effects on the observables of SGL for disc galaxies, e.g., SGL cross section, within the MOND scenario. We draw comparison to known results on disc galaxy lensing in $\Lambda$CDM, and we show that: \textbf{(i)} the number count of disc galaxy lenses predicted by MOND exceeds the one expected in the DM scenario, and \textbf{(ii)} in MOND disc galaxy lensing the inclination effects are enhanced w.r.t.\ the DM halo paradigm. With the upcoming optical surveys, LSST, DES and Euclid, these differences in the inclination effects and overall lens number counts will become statistically detectable in SGL of disc galaxies, as these surveys are predicted to observe hundreds of thousands of galaxy-scale lenses \citep{Collett_2015,Barroso_2024}. Thus, we propose that, in the next decade, statistical studies on the inclination and number of disc galaxy lenses will lead to an important direct test to differentiate cold DM from MOND.

Here, we focus on disc galaxies as isolated systems, assuming no interfering matter along the line of sight. This is a typical practice in studies of SGL on galaxy scales, where matter contributing to the lens effect from outside the lens galaxy is modelled by fitting an external convergence parameter, often in combination with weak lensing measurements \citep[see e.g.,][]{Tihhonova_2018,Wells_2023}. However, we note that due to external field effects (EFE) in the MOND paradigm the assumption of an isolated system is ultimately a stronger requirement than in the DM scenario \citep{Famaey_2012, Banik_2021}. Indeed, due to the non-linear nature of the MOND field equation (cf., Eqs.\ \eqref{eq:AQUAL} and \eqref{eq:QMOND}), the presence of external matter non-trivially influences the internal gravitational field of the lens itself. Therefore, our study will be directly applicable to highly isolated disc galaxies, whilst case-by-case corrections for EFE will have to be considered for SGL within clusters and more nuanced environments.

Throughout this paper, we assume a spatially flat $\Lambda$CDM cosmology, with matter density $\Omega_{\text{M}0}=0.3$ and Hubble constant $H_0 = 70\,$km\,s$^{-1}$Mpc$^{-1}$, as a background cosmological model, with lens and source placed at redshift $z_{\text{L}} = 0.5$ and $z_{\text{S}} = 2.0$, respectively. We note that assuming a spatially flat $\Lambda$CDM cosmology is not at odds with a MOND description of disc galaxies. Indeed, even if MOND were to be ultimately correct, its relativistic extension would have to closely match the cosmological evolution predicted by the highly tested $\Lambda$CDM model. Therefore, employing $\Lambda$CDM as a cosmological model within a MOND scenario at the galactic scales can be directly understood as a good first-order approximation to the possible choice of a correct, and yet unknown, MOND cosmology \citep{Famaey_2012,Banik_2021}. 

The rest of the paper is structured as follows. In Section \ref{sec:SGL} we recall the SGL fundamentals and derive the observables used to probe inclination effects in disc galaxy SGL. In Section \ref{sec:DenProf} we define the analytic baryon density distribution for the disc galaxy models, and we derive the respective PDM distribution for realistic templates of such galaxies. In Section \ref{sec:Results} we report our results, focusing on the effect of disc galaxy inclination on SGL cross sections and draw comparison to the DM scenario. Finally, Section \ref{sec:Conclusions} is dedicated to conclusions and future perspectives.

\section{Strong gravitational lensing}\label{sec:SGL}
\begin{figure}
\label{FigureLensing}
		\centering
		\includegraphics[width=0.49\textwidth]{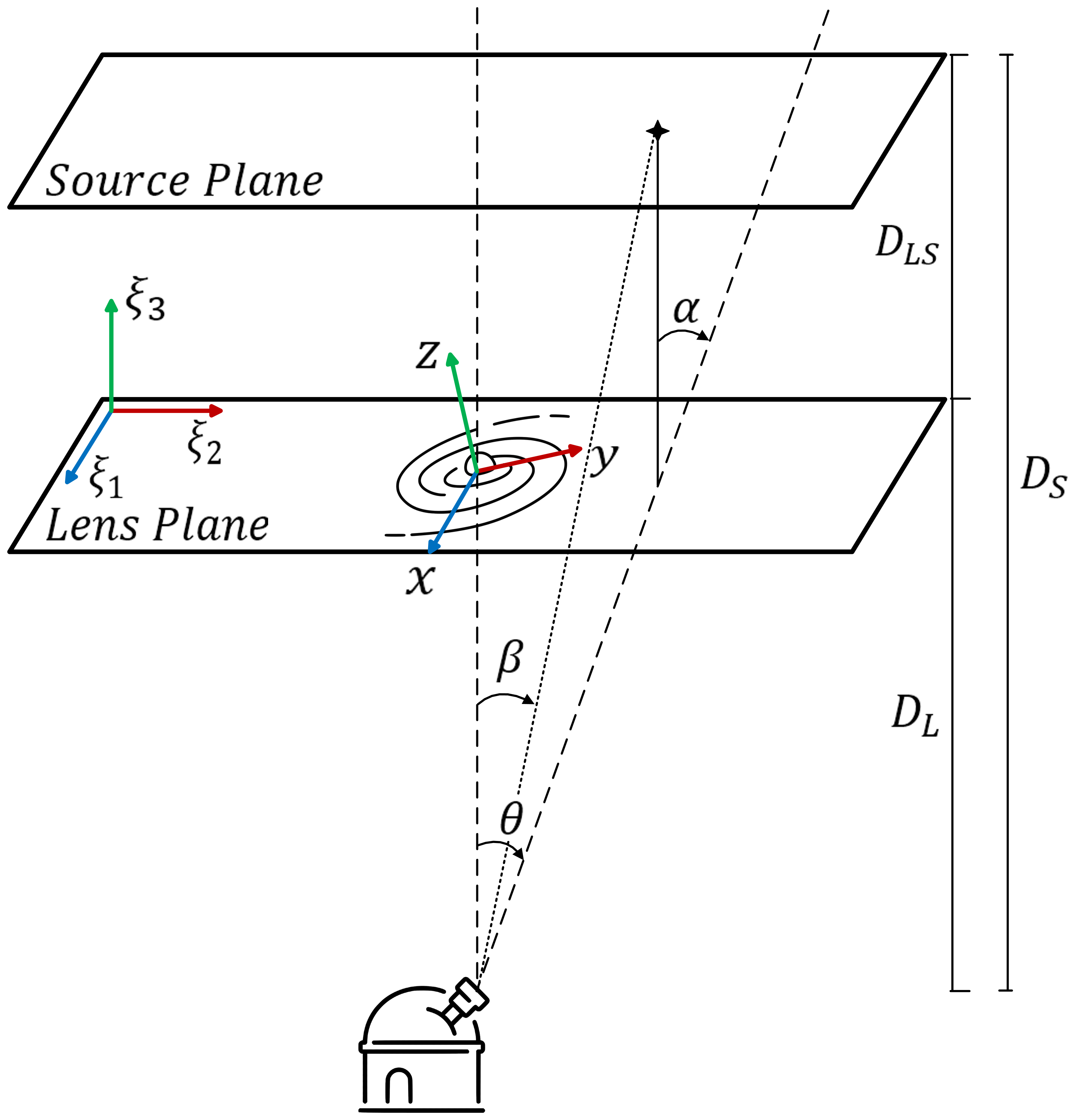}
		\caption{Lensing configuration for SGL by a disc galaxy. We report source, lens and observer positions, as well as the relevant angles and coordinate systems. Here, $\alpha$ is the deflection angle, $\beta$ is the angular source position and  $\theta$ is the angular image position.}
            \label{fig:lensing_geometry}
\end{figure}

For a given density distribution, in the weak-field limit of gravitational fields, the deflection angle of a light ray propagating perpendicular to the plane of the lens, $\vec{\hat{\alpha}}$, is defined as\footnote{Employing \eqref{eq:alpha} implies a restriction on the possible choice of the a priori unknown relativistic MOND extension. Indeed, we are assuming that the lensing potential is twice the Newtonian potential in the weak-field regime, as in general relativity (see e.g., \cite{Bekestein_2004} for such a theory).}
\begin{equation}
    \label{eq:alpha}
    \vec{\hat{\alpha}}(\xi_1,\xi_2) := \frac{2}{c^2} \int^\infty_{-\infty} \text{d}\xi_3 \,  \vec{a}_\perp(\vec{\xi}) \, .
\end{equation}
Here, $\vec{a}_\perp$ denotes the acceleration due to the gravitational potential of the lens object along the lens plane, i.e.,
\begin{equation}
    \vec{a}_\perp = - \vec{\nabla}_\perp \PhiN \, ,
    \label{eq:acc}
\end{equation}
with
\begin{equation}
    \label{eq:Nprop}
    \PhiN(\vec{\xi}) = G \iiint \frac{ \rho(\vec{r})}{|\vec{\xi}-\vec{r}|} \, \text{d}^3 \vec{r} \,.
\end{equation}
It is important to stress that the lens plane lies orthogonal to the line of sight from the observer to the lens and thus, it is not dependent on the inclination of the lens itself (see Fig.\ \ref{fig:lensing_geometry}). Therefore, the inclination of a disc galaxy will have a large impact on its lensing, which is derived from the projection of its baryonic matter content onto the lens plane. Here, we note that in the presence of (almost) spherical DM halos, the inclination effects of disc lenses, although still clearly present, are partially suppressed by the role played by the DM in the lensing phenomena \citep{Keeton_1998a}. However, we do not expect such a situation in MOND, where no DM halos are present and the lensing is driven solely by the baryonic disc.

We note that by employing Eqns \eqref{eq:alpha}-\eqref{eq:Nprop} coupled to the ENS of the disc galaxy's gravitational potential, SGL in MOND is then easily integrated into the usual lensing formalism, with $\PhiM$ in place of the classical Newtonian potential $\PhiN$.

We can now follow the standard lensing formalism \citep[see e.g.,][]{Schneider_2006,Meneghetti_book_2022} to define the lens equation
\begin{equation}\label{eq:lens_eq}
    \vec{\beta} = \vec{\theta} - \frac{D_{LS}}{D_S}\vec{\hat{\alpha}} \, (\vec{\theta}) = \vec{\theta} - \vec{{\alpha}}(\vec{\theta}) \, ,
\end{equation}
where $\vec{\beta}$ and $\vec{\theta}$, respectively, describe the angular position w.r.t.\ to the lens centre of the source object and of the lensed images. Moreover, $D_{LS}$ and $D_S$ represent the angular diameter distances from the lens to the source and from the observer to the source. Fig.\ \ref{fig:lensing_geometry} depicts the lensing configuration under study. 
Then, from \eqref{eq:lens_eq} we can define the Jacobian matrix $\mathcal{A}$,
\begin{equation}
    \mathcal{A} =  \frac{\partial \vec{\beta}}{\partial \vec{\theta}} = \boldsymbol{1} - \grad\vec{{\alpha}} \,.
\end{equation}
The critical curves for the lens configuration are defined as the positions where the magnification, $\mu := 1 / \det \mathcal{A}$, is formally infinite (for a point source). We note that it is, however, generally convenient to determine critical curves directly from the condition $\det \mathcal{A} = 0$. Then, we can use the lens equation \eqref{eq:lens_eq} to project the critical curves back onto caustic curves on the source plane.  

The position of a source object with respect to the caustic lines determines the number of images produced and their respective positions. Objects within caustic lines will produce multiple images, and thus, the area enclosed by caustic lines on the source plane defines the lensing cross section $\sigma$ \citep{Schneider_2006,Meneghetti_book_2022}. We note that whilst the lensing cross section is a survey-independent quantity, the likelihood of observing SGL is not. The latter, although directly proportional to the former, will be influenced by the specific observations carried out \citep{Collett_2015,Meneghetti_2022}. Therefore, to characterise the inclination effects of disc galaxies in MOND we only consider $\sigma$ as an indirect measure of SGL likelihood and do not specialise to any specific surveys.

Finally, in this paper, we focus on point sources. Extended sources will have higher lensing likelihood, but for the purposes of determining the differences between lensing effects for different disc galaxy mass profiles and inclination effects, considering point-like sources is sufficient.

\section{Matter density profiles}\label{sec:DenProf}

To describe the baryonic component of disc galaxies,  we couple a thick exponential disc to a spherical bulge, such that $\rhoB = \rhob + \rhod$, where $\rhob$ and $\rhod$ are the densities of the bulge and the disc, respectively.

\subsection{Analytical baryonic matter density modelling}
The spherical bulge component is well described by the Plummer density profile \citep{Pouliasis_milky_2017}
\begin{equation}\label{eq:rhob}
    \rhob (r) = \frac{3\rb^2\Mb}{4\pi\left(r^2+\rb^2\right)^{5/2}} \,,
\end{equation}
where $r$ is the spherical radius, $\rb$ is the scale parameter of the bulge and $\Mb$ is the total bulge mass. The baryonic disc matter distribution is conventionally taken as $\rhod(R, z)=\Sigma(R)\,Z(z)$, where $R$ is the cylindrical radius and $z$ is the axial coordinate. Then, the surface density of the disc is well approximated by the exponential profile \citep[see e.g.,][]{Binney_2008}
\begin{equation}
\label{eq:rhodR}
    \Sigma(R)=\frac{\Md}{2\pi \Rd^2}e^{-R/\Rd}\, ,
\end{equation}
where $\Md$ is the total disc mass and $\Rd$ is the scale length in the radial direction for the galactic disc. The distribution along $z$, such that $\int Z(z)\, \text{d}z=1$, can be defined as \citep{Binney_2008}
\begin{equation}
\label{eq:rhodZ}
    Z(z)=\frac{1}{2\zd}e^{-|z|/\zd} \, ,
\end{equation}
where $\zd$ is the thickness scale length of the disc galaxy. In separating the density profile in this way, we have assumed that the disc has a scale height that is constant with radius, as is observed in the photometry of local edge-on disc galaxies \citep{Van_1981,Van_1982}. Thus, we have
\begin{equation}
\label{eq:rhod}
    \rhod(R,z) = \frac{\Md}{2\pi\Rd^2\zd}e^{-R/\Rd}e^{-|z|/\zd} \, .
\end{equation}
We recall that in the pioneering work of \cite{Keeton_1998a} on SGL by disc galaxies within the $\Lambda$CDM paradigm, the baryonic disc component was approximated either by Mestel \citep{Mestel_1963} or Kuzmin discs \citep{Kuzmin_1956} within a softened, oblate isothermal DM halo\footnote{We note that in the papers by \cite{Wang_1997} and \cite{Moeller_1998}, uniform discs were considered for modelling the lens effect of a disc galaxy. However, due to the uniform assumption, and as noted by \cite{Keeton_1998a}, these cannot produce realistic rotation curves and do not depict actual three-dimensional density profiles of disc galaxies.}. Moreover, similar chameleon profiles were employed to combine dynamical and SGL data for disc galaxies in the more recent studies by \cite{Dutton_2011} and \cite{Suyu_2012}, as well as in the first consideration of non-spherical lenses in MOND \citep{Shan_2008}. These profiles have been preferred to fully exponential mass distributions for the disc component due to their analytical properties in SGL computation \citep{Keeton_1998a, Shan_2008, Dutton_2011, Suyu_2012}. 

In MOND the lensing is entirely defined by the baryonic components of the galaxies, as no DM is required. Thus the morphological characteristics of the baryons in disc galaxies are more important in MOND than for cold DM scenarios. Furthermore, as already discussed by \cite{Keeton_1998a}, Mestel and Kuzmin discs represent a poor fit for the vertical component of the baryonic disc and can only be a first approximation for the exponential radial profile. Hence, to avoid biases in the inclination effects in MOND, we chose to maintain a closer match to observational data for the mass density profile modelling of disc galaxies and implement a realistic exponential disc profile coupled to a spherical bulge.

\subsection{Realistic disc galaxy model parameters}
Once the baryonic mass profile has been defined, we have to pick meaningful values for the parameters so that the resulting disc galaxy models can be understood to be a good representation of observational data. 

Following \cite{Keeton_1998a} we consider a massive spiral galaxy with a maximal disc, such that DM (or MOND) effects play a major role only in the outer kinematics of the galaxy. We therefore select a massive disc galaxy with $\Md = 10^{11}\,$M$_\odot$, and $\Rd = 3.5 \, \text{kpc}$. We then consider three possible values for $\zd$, namely $\zd \in \{0.35, 0.105, 0.035\}$ kpc, so that the ratio $\zd/\Rd \in \{0.1,0.03,0.01\}$ as chosen by \cite{Keeton_1998a}, thus facilitating comparisons. 

We must also select appropriate values for $\Mb$ and $\rb$. We note that correlations between $\Mb$, $\rb$ and the disc parameters have been established in the literature \citep[see, e.g.,][]{Lang_2014,Allen_2017,Forbes_2018}. However, these are better understood for low-mass spirals and, given the plethora of disc galaxy morphologies, these correlations produce only weak constraints on the bulge parameters once the disc has been selected. Therefore, in our analysis, we selected $\Mb/\Md \in \{0.01,0.1,0.2\}$ and $\rb/\Rd \in \{0.03,0.1,0.2\}$.

Finally, we point out that our choice of following \cite{Keeton_1998a} in studying SGL by disc galaxies only in massive spirals is well justified by noting that the likelihood of SGL increases with the mass. Therefore, the low-mass spirals are not expected to significantly contribute to the population of disc galaxy lenses, even in the upcoming optical surveys.

\subsection{Equivalent Newtonian system}\label{sec:ENS}

To build the ENS corresponding to our choice of $\rhoB = \rhod +\rhob$ we exploit the linearity of Poisson equation, so that
\begin{equation}\label{eq:potentials}
    \PhiN = \Phi_\text{N,d} + \Phi_\text{N,b} \, ,
\end{equation}
where $\Phi_\text{N,d}$ and $\Phi_\text{N,b}$ are the Newtonian gravitational potentials generated by the exponential disc and the spherical bulge, respectively. We can directly calculate the two potentials by solving Eq.\ \eqref{eq:Poisson} with \eqref{eq:rhob} and \eqref{eq:rhod} as sources. We then obtain
\begin{align}
    \Phi_{\text{N,d}}(R,z) = - G \Md  \int^\infty_0 \text{d}k \frac{J_0 (R k) \left(e^{-|z| k} - \zd k e^{-|z|/\zd}\right)}{\left( 1 + \Rd^2 \, k^2 \right)^{3/2}\left(1 - \zd^2 k^2\right)}  \, , \label{eq:PhiNd}
\end{align}
\begin{equation}
    \Phi_{\text{N,b}}(r) = \frac{G\Mb}{\rb} \left( 1 - \frac{\rb}{\sqrt{\rb^2 + r^2}} 
     \right) \, , \label{eq:PhiNb}
\end{equation}
where $J_0$ is the Bessel function of the first kind. We can now directly calculate $\grad\PhiN$ and exploit Eq.~\eqref{eq:PDM} to obtain $\PDM$ once we have chosen a functional form for the interpolating function $\tilde{\nu}$. In this work, we implement the same functional form used in the Phantom of Ramses code for N-body and hydrodynamical simulations in the MOND scenario\footnote{Here, we have implemented an original routine to derive the PDM distribution and the SGL effect of the disc galaxies. The codes implemented are available at \href{https://github.com/chrisharhaw/MOND_lensing.git}{https://github.com/chrisharhaw/MOND\_lensing.git}. Moreover, we give a description of these codes in Appendix A.}\citep{Candlish_2014,Lüghausen_2015}, i.e.,
\begin{equation}
    \label{eq:our_nu}
    \tilde{\nu}(y) = -\frac{1}{2} + \sqrt{\frac{1}{4} + \frac{1}{y}}\, .
\end{equation}
Here, it is important to note that whilst studies of disc galaxy rotation curves are quite robust w.r.t.\ the choice of interpolating function (due to the onset of the dMOND regime), in SGL this choice is highly non-trivial. Indeed, SGL effects probe the full spectrum of applicability of MOND and thus will be affected by the choice of $\tilde{\nu}$\footnote{To test the dependence of our results on the choice of interpolating function, in Appendix B we show a comparison of our main results with those obtained with the RAR interpolating function \citep[see e.g.,][]{McGaugh_2016,Stiskalek_2023}. We find only marginal differences w.r.t.\ the results obtained for $\tilde{\nu}$ given in Eq.\ \eqref{eq:our_nu}.}. In this paper, we have chosen the widely adopted form of interpolating function of Eqn.\ \eqref{eq:our_nu} as it has been proven to give a MOND realisation able to pass a variety of observational tests \citep{Famaey_2012}, and it is widely employed in the study of in-galaxy dynamics within the MOND scenario \citep[see e.g.,][]{Re_2023, DiCintio_2024}. However, we caution that the results may change with different choices of the interpolating function, as already noted in previous SGL MOND studies \citep[see, e.g.,][]{Sanders_2008}.

\begin{figure}[htbp]
    \centering
    \begin{minipage}[b]{0.5\textwidth}
        \includegraphics[width=\linewidth]{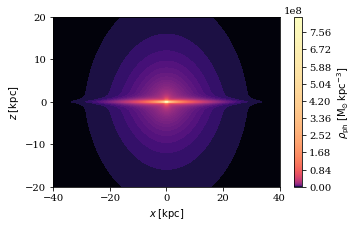}
    \end{minipage}
    \hspace{0.5cm}
    \begin{minipage}[b]{0.5\textwidth}
        \includegraphics[width=\linewidth]{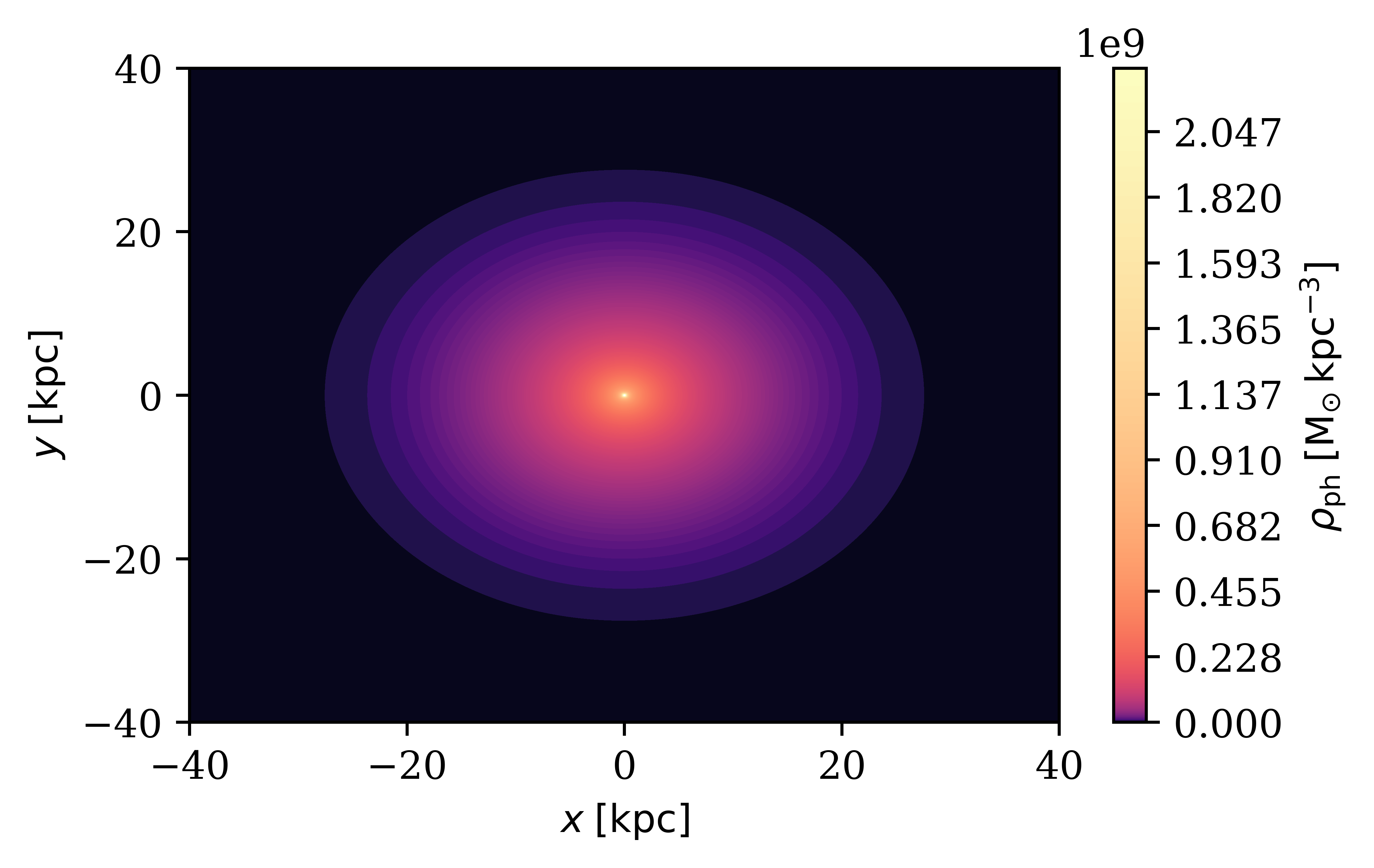}
    \end{minipage}
    \caption{PDM density distribution on the $x-z$ plane (upper panel) and on the $x-y$ plane (lower panel) for a disc galaxy with $\Md = 10^{11}\, \text{M}_\odot$, $\Rd = 3.5\, \text{kpc}$, $\zd = 0.35\, \text{kpc}$, $\Mb = 10^{9}\, \text{M}_\odot$, and $\rb = 0.7\, \text{kpc}$. A log-scale is used for the density to highlight the morphology of the PDM distribution.}
    \label{fig:rho_density}
\end{figure}

\begin{figure}[htbp]
    \centering
    \begin{minipage}[b]{0.5\textwidth}
        \includegraphics[width=0.9\linewidth]{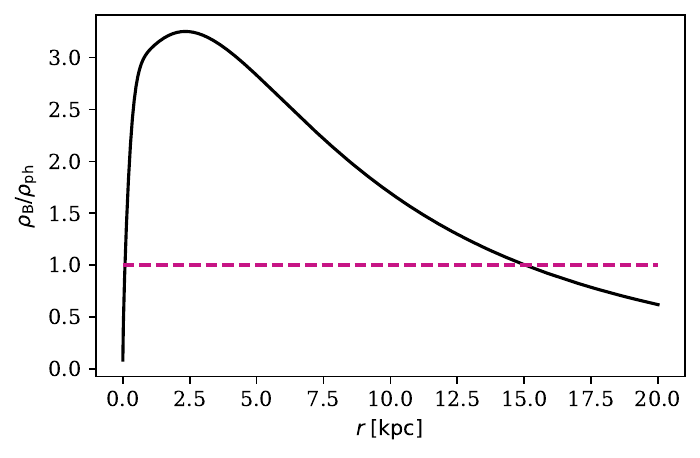}
        \label{fig:PDM_overtakeA}
    \end{minipage}
    \hspace{0.5cm}
    \begin{minipage}[b]{0.5\textwidth}
        \includegraphics[width=0.95\linewidth]{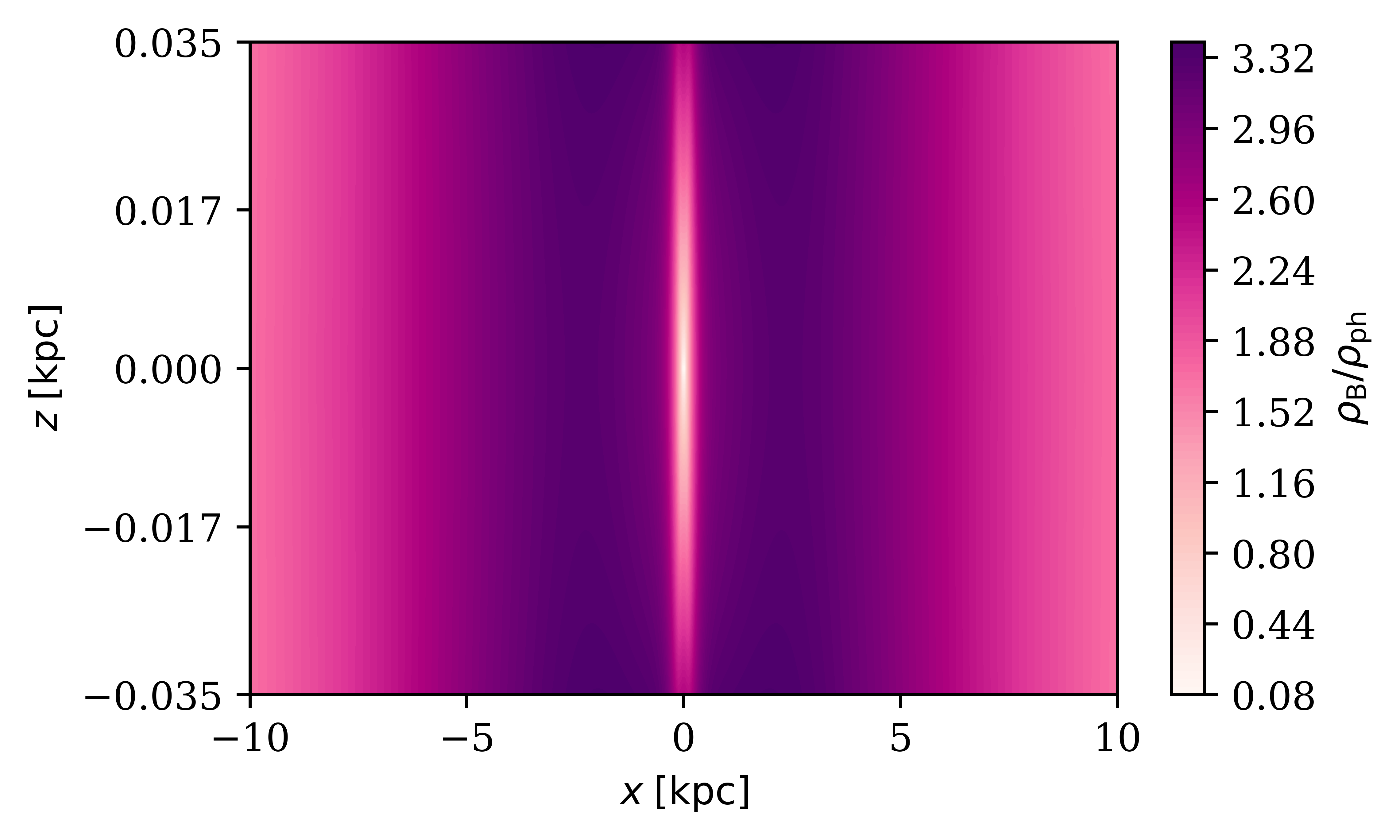}
        \label{fig:PDM_overtakeB}
    \end{minipage}
    \caption{Ratio of the baryon density to the PDM density, $\rho_\text{B}/\PDM$, along the radial direction on the equatorial plane (top panel), and the corresponding two-dimensional distribution in the $x-z$ plane within a scale height of the galactic disc (bottom panel) for a disc galaxy with $\Md = 10^{11}\, \text{M}_\odot$, $\Rd = 3.5\, \text{kpc}$, $\zd = 0.35\, \text{kpc}$, $\Mb = 10^{9}\, \text{M}_\odot$, and $\rb = 0.7\, \text{kpc}$. The purple dashed line in the top panel highlights the point at which the two densities have the same value.}
    \label{fig:PDM_overtake}
\end{figure}

In Fig.\ \ref{fig:rho_density} we show as an example, the distribution of PDM from an edge-on and face-on point of view for a disc galaxy with $\Md = 10^{11}\, \text{M}_\odot$, $\Rd = 3.5\, \text{kpc}$, $\zd = 0.35\, \text{kpc}$, $\Mb = 10^{9}\, \text{M}_\odot$, and $\rb = 0.7\, \text{kpc}$. For the same galaxy, in Fig.\ \ref{fig:PDM_overtake} we show the ratio of the baryon density to the PDM density, $\rho_\text{B}/\PDM$, along the radial direction on the equatorial plane, and the corresponding two-dimensional distribution in the $x-z$ plane within a scale height of the galactic disc. From Fig.\ \ref{fig:rho_density} we note that the resulting PDM distribution follows the same symmetries of the underlying baryonic matter, as we should expect from Eqn.\ \eqref{eq:PDM}. We find it to be always positive definite and with no peak shift in its concentration w.r.t.\ the baryons. Moreover, we recover the superimposed disc-halo PDM morphology shown in previous studies \citep[see e.g.,][]{Lüghausen_2015}.  

Finally, from Fig.\ \ref{fig:PDM_overtake} we see that the PDM overtakes the baryonic matter as the dominant ``matter'', component within the very core of the galaxy (below the parsec level) and at large distances from the galactic centre (i.e., approximately at $15$ kpc), precisely where we would expect MOND (or DM) effects to dominate the dynamics for this type of galaxy. The overtake of the PDM over the baryons within the centre of the disc galaxy indicates the presence of a cusp in the PDM distribution. We note that such behaviour is not related to the specific choice of interpolating function, as the centre of the galaxy is in a dMOND regime, but relates to the interplay of the latter with the baryon density profile. As such, it is a feature shared by any interpolating function which correctly encodes the dMOND limit.

\section{Results}\label{sec:Results}

\begin{figure*}
    \begin{minipage}{0.5\textwidth}
        \includegraphics[width=\linewidth]{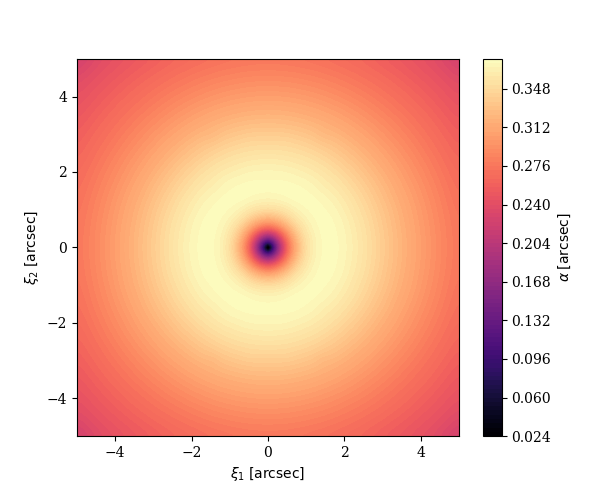}
        \centering
        (a) Deflection angle, $i = 0$.
    \end{minipage}%
    \begin{minipage}{0.5\textwidth}
        \includegraphics[width=\linewidth]{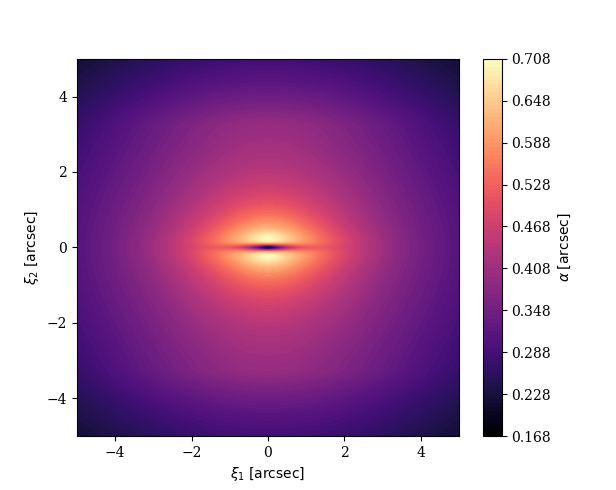}
         \centering
         (b) Deflection angle, $i = \pi/2 $.
    \end{minipage} \\
    \vspace{0.01cm}
    \begin{minipage}{0.5\textwidth}
        \includegraphics[width=\linewidth]{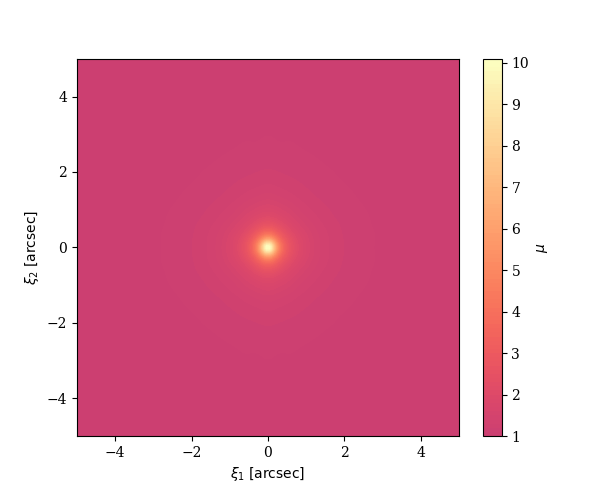}
         \centering
         (c) Magnification, $i = 0$.
    \end{minipage}%
    \begin{minipage}{0.5\textwidth}
        \includegraphics[width=\linewidth]{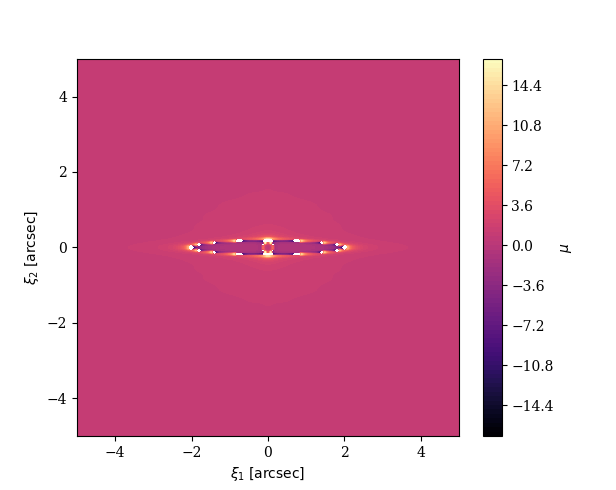}
         \centering
         (d) Magnification, $i = \pi/2 $.
    \end{minipage} \\
    \vspace{0.01cm}
    \begin{minipage}{0.5\textwidth}
        \includegraphics[width=\linewidth]{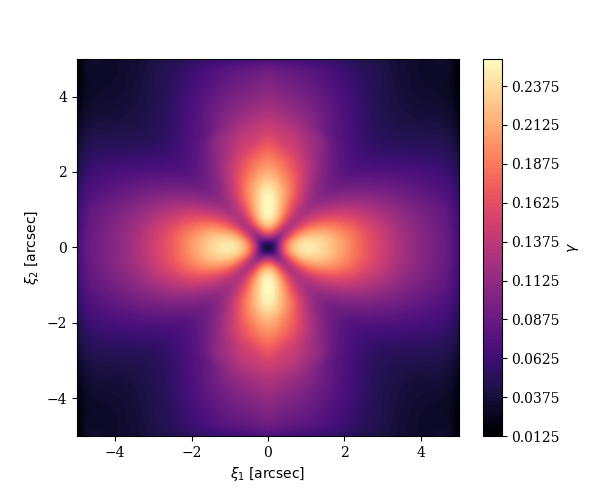}
         \centering
         (e) Shear, $i = 0 $.
    \end{minipage}%
    \begin{minipage}{0.5\textwidth}
        \includegraphics[width=\linewidth]{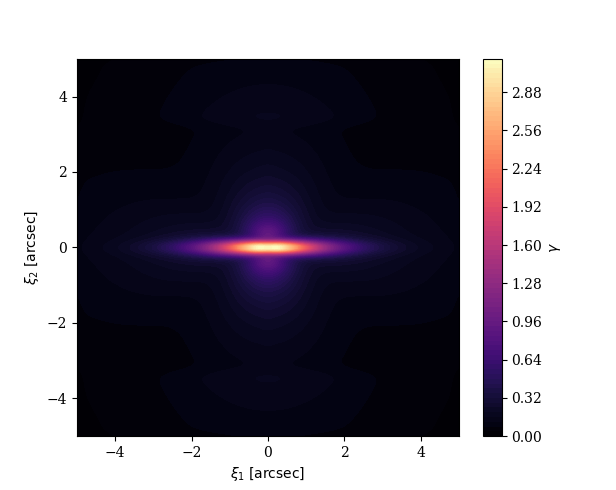}
         \centering
         (f) Shear, $i = \pi/2 $.
    \end{minipage}
    \caption{Example of lensing observables, namely the deflection angle ($\alpha$), the magnification ($\mu$), and the shear ($\gamma$), for a disc galaxy of inclinations $i = 0$ (left) and $i = \pi/2$ (right). The disc galaxy model is obtained by fixing $\Md = 10^{11}\, \text{M}_\odot$, $\Rd = 3.5\, \text{kpc}$, $\zd = 0.35\, \text{kpc}$, $\Mb = 10^{9}\, \text{M}_\odot$, and $\rb = 0.7\, \text{kpc}$.}
    \label{fig:magnitude_magnification_shear}
\end{figure*}

In Fig.\ \ref{fig:magnitude_magnification_shear} we show an example of the deflection angle magnitude, $\alpha$, the magnification, $\mu$, and the shear, $\gamma$, for the example of a massive disc galaxy with $\Md = 10^{11}\, \text{M}_\odot$, $\Rd = 3.5\, \text{kpc}$, $\zd = 0.35\, \text{kpc}$, $\Mb = 10^{9}\, \text{M}_\odot$, $\rb = 0.7\, \text{kpc}$ and $i = 0^\circ,\,90^\circ$. As expected, in the face-on case the geometry of the lensing agrees with that of an axisymmetric lens, and moreover gives physically plausible deflection angles. This is consistent throughout the array of inclination angles, as can be seen for the edge-on case, in which the maximum value of $\alpha$ approximately doubles w.r.t.\ to the face-on scenario. For $i = 90^\circ$, the breaking of the circular symmetry of the projected matter density on the lens plane drastically changes the geometry of the lensing effects, already showing the importance of inclination for disc galaxy lensing in MOND. 

\begin{figure*}
    \centering
    \includegraphics[width=\linewidth]{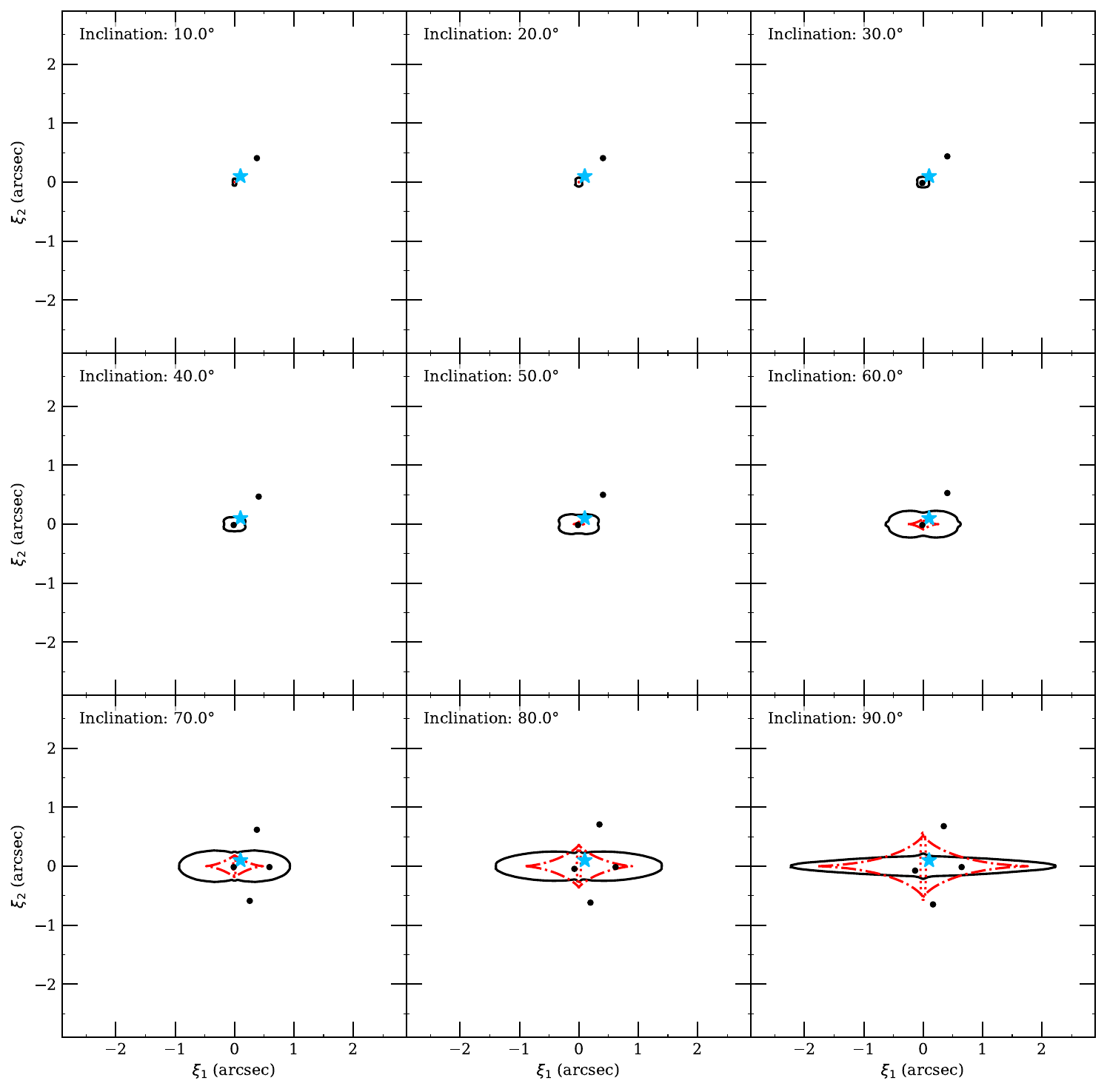}
    \caption{Example image configurations, caustic curves and critical curves for spiral galaxy MOND lenses with varying inclination and point source objects placed at a constant position relative to the lens centre. The solid black line, dotted red line and dashed-dotted red-line indicate the critical curve, radial caustic curve and tangential caustic curves respectively. The blue star represent the source image and black dots show the corresponding lensed images. The parameters of the lens galaxy are fixed to be $\Md = 10^{11}\, \text{M}_\odot$, $\Rd = 3.5\, \text{kpc}$, $\zd = 0.35\, \text{kpc}$, $\Mb = 10^{9}\, \text{M}_\odot$, and $\rb = 0.7\, \text{kpc}$. The lens and source are placed at redshift $z_{\text{L}} = 0.5$ and $z_{\text{S}} = 2.0$, respectively.}
    \label{fig:3by3caustics_inc}
\end{figure*}

We show in Fig.\ \ref{fig:3by3caustics_inc}, for a fixed source position and lens parameters, that a wide variety of image configurations can be obtained by changing the inclination angle of the lens w.r.t.\ the observer. Indeed, we recover familiar image configurations within our results -- such as the Einstein cross quad-image configuration -- already found in the previous studies on similar lensing systems. Here, we note that the tangential caustic curve of the lens do not develop butterfly features, even at high inclinations. This is in agreement with the results of \cite{Keeton_1998a} for disc galaxies within the DM paradigm, and in contrast with the findings of \cite{Wang_1997} and \cite{Moeller_1998}. However, the density modelling employed by these studies presents unphysical features which might then influence the morphology of the tangential caustic curve (see \cite{Gilles_2009} for a discussion on exotic caustic curve morphologies).

\begin{figure*}[h!]
    \centering
    \includegraphics[width=0.8\textwidth]{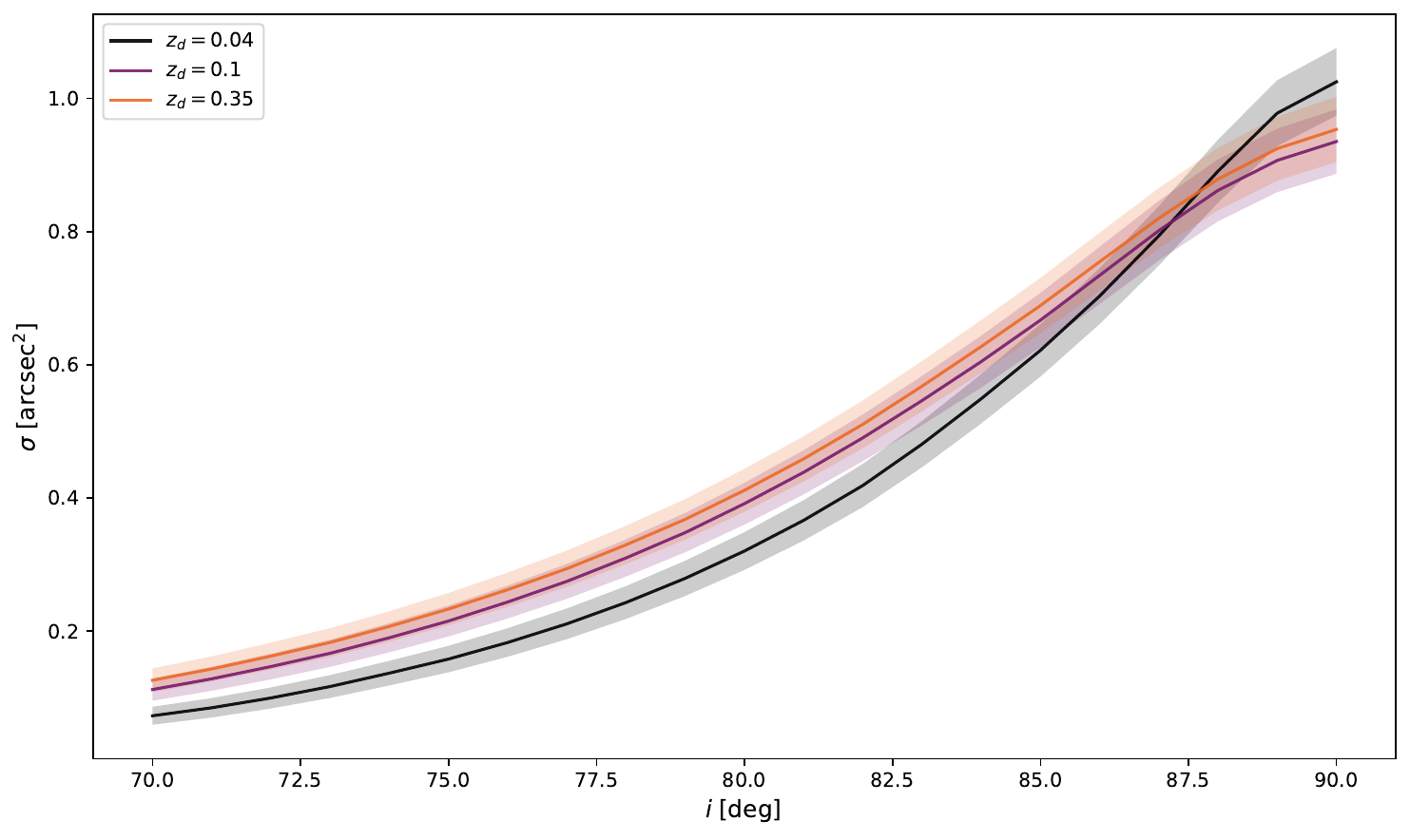}
    \caption{Impact of disc thickness $\zd$ on lensing cross section $\sigma$ with varying inclination angle within the MOND scenario. The other corresponding lens galaxy parameters are $\Md = 10^{11}\, \text{M}_\odot$, $\Rd = 3.5\, \text{kpc}$,  $\Mb = 10^{9}\, \text{M}_\odot$, and $\rb = 0.7\, \text{kpc}$. The lens and source are placed at redshift $z_{\text{L}} = 0.5$ and $z_{\text{S}} = 2.0$, respectively. The shaded areas represent the one standard deviation uncertainties around the obtained SGL cross section curves.}
    \label{fig:zd_inc}
\end{figure*}

\begin{figure*}
    \centering
    \includegraphics[width=0.8\textwidth]{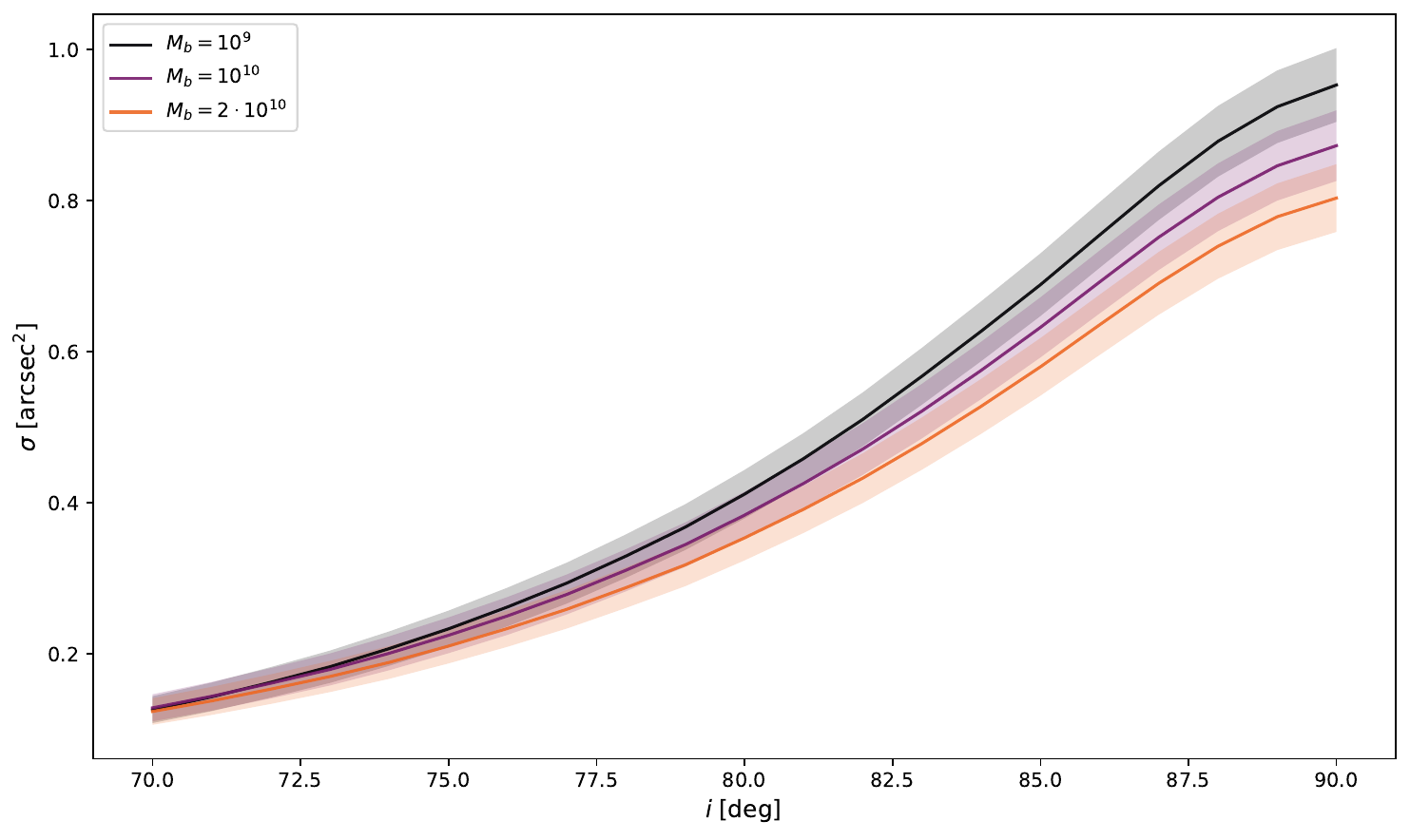}
    \caption{Impact of bulge mass $\Mb$ on lensing cross section $\sigma$ with varying inclination angle within the MOND scenario. The other corresponding lens galaxy parameters are $\Md = 10^{11}\, \text{M}_\odot$, $\Rd = 3.5\, \text{kpc}$,  $\zd = 0.35\, \text{kpc}$, and $\rb = 0.7\, \text{kpc}$. The lens and source are placed at redshift $z_{\text{L}} = 0.5$ and $z_{\text{S}} = 2.0$, respectively. The shaded areas represent the one standard deviation uncertainties around the obtained SGL cross section curves.}
    \label{fig:Mb_inc}
\end{figure*}

We find that changing the lens parameters, restricted to realistic disc galaxy models, does not significantly influence the shape of critical or caustic curves produced. However, varying the lens morphology and mass distribution in this way led to direct changes in the overall caustic area, as is shown in Fig.~\ref{fig:zd_inc} and Fig.~\ref{fig:Mb_inc}.

\begin{figure*}
    \centering
    \includegraphics[width=0.8\textwidth]{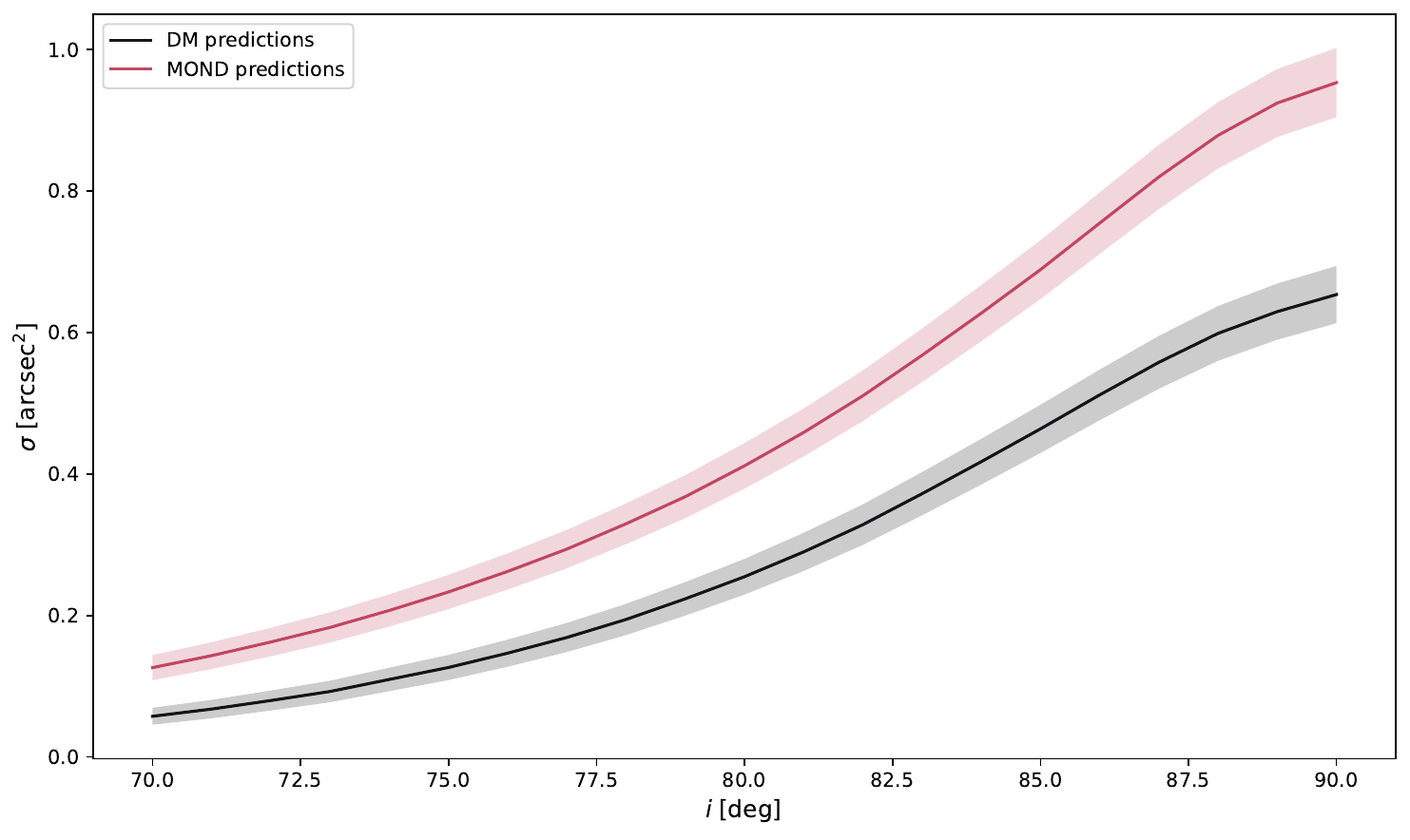}
    \caption{A comparison of lensing cross sections predictions with varying inclination angle for the same disc galaxy in MOND and the conventional DM scenario. The disc galaxy parameters are $\Md = 10^{11}\, \text{M}_\odot$, $\Rd = 3.5\, \text{kpc}$,  $\zd = 0.35\, \text{kpc}$, and $\Mb = 10^{9}\, \text{M}_\odot$, and $\rb = 0.7\, \text{kpc}$. The DM halo is taken as a NSIS profile with parameters $\rho_0 = 6.36\cdot 10^7\, \text{M}_\odot/\text{kpc}^3$, and $r_0 = 3.44$ kpc. The lens and source are placed at redshift $z_{\text{L}} = 0.5$ and $z_{\text{S}} = 2.0$, respectively. The shaded areas represent the one standard deviation uncertainties around the obtained SGL cross section curves.}
    \label{fig:MONDvDM_inc}
\end{figure*}

In Figs. \ref{fig:zd_inc}-\ref{fig:MONDvDM_inc} we show the lensing cross section with varying inclination angle and an ensemble of disc galaxy parameters. In these plots, due to the nature of calculating the SGL cross section via numerical methods on a grid, we assume a Poissonian error over the pixel count contained within the tangential caustic curves, i.e., the square root of the number of pixels (multiplied by the single-pixel area value).  

Fig.\ \ref{fig:zd_inc} shows the effects of varying the disc thickness $\zd$ on the lensing cross section. Interestingly, no clear trend w.r.t.\ the disc thickness emerges, already indicating the non-trivial interplay between non-linear MONDian effects and lens parameters. Indeed, an increased lensing likelihood for more diffuse disc masses, is in contrast to the standard lensing expectations. In the weak-field limit of general relativity, a more concentrated mass would produce stronger lensing effects.

In Fig.~\ref{fig:Mb_inc} the non-linear, non-Newtonian effects of MOND are more clearly captured. We see that an increased bulge mass negatively correlates with SGL cross section, within the mass range explored. Such a behaviour is in stark contrast to expectations from the standard weak-field SGL of disc galaxy \citep{Keeton_1998a}. This trend can be understood by noting that SGL encompasses the full breadth of MOND effects, as it probes the Newtonian limit, the dMOND regime and, most importantly, the transition acceleration scales. Therefore, an increased mass in the bulge, whilst increasing the Newtonian lensing effects, it is found to decrease the MONDian contribution to lensing (which is ultimately dominant in this case). Indeed, a greater mass induces larger Newtonian acceleration and hence decreases the onset of MONDian effects.

\begin{figure*}
    \centering
    \includegraphics[width=0.8\textwidth]{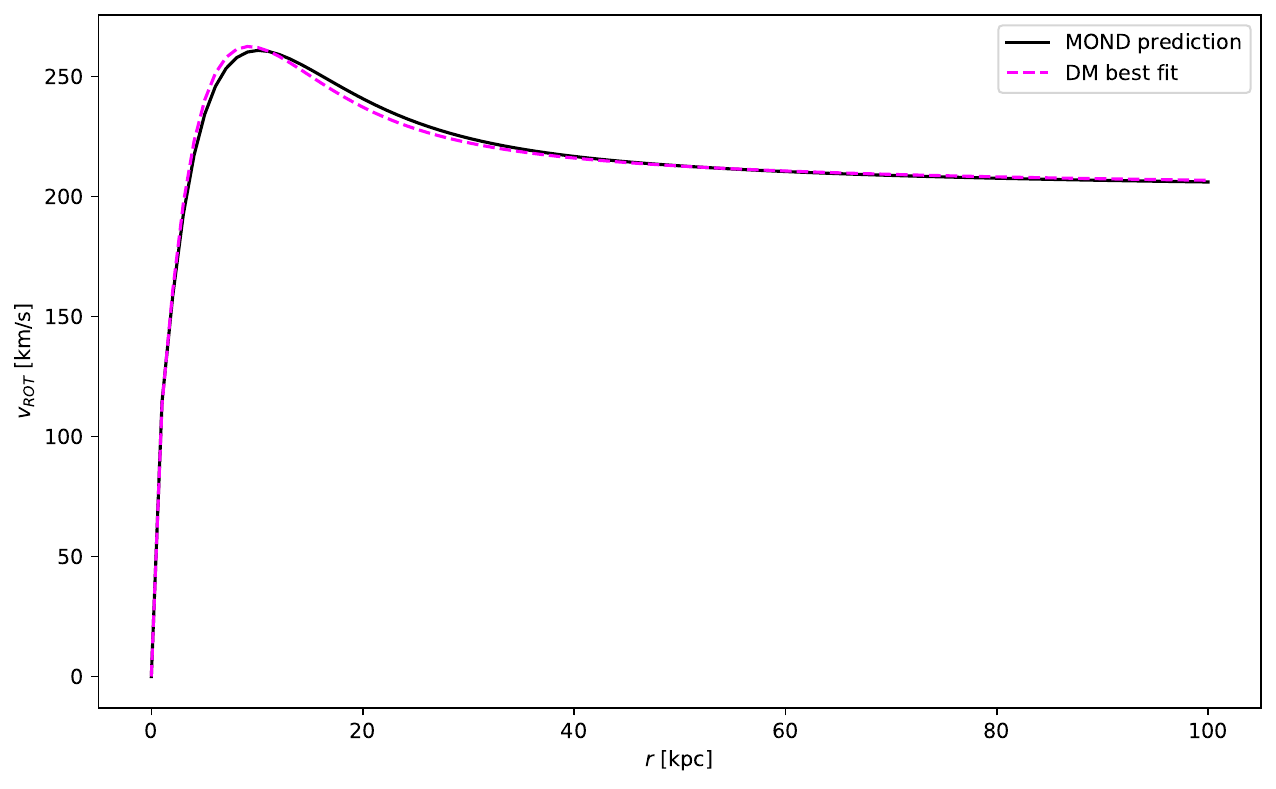}
    \caption{A comparison of MONDian rotation curve predictions with the best fit DM alternative. The solid blue line shows the MONDian rotation curve prediction for a disc galaxy with parameters: $\Md = 10^{11}\, \text{M}_\odot$, $\Rd = 3.5\, \text{kpc}$,  $\zd = 0.35\, \text{kpc}$, and $\Mb = 10^{9}\, \text{M}_\odot$, and $\rb = 0.7\, \text{kpc}$. The dashed red line show the corresponding Newtonian fit of a DM halo, which is taken as a NSIS profile with parameters $\rho_0 = 6.36\cdot 10^7\, \text{M}_\odot/\text{kpc}^3$, and $r_0 = 3.44$ kpc}. 
    \label{fig:rot_curve}
\end{figure*}

Finally, in Fig.\ \ref{fig:MONDvDM_inc} we present a comparison between the SGL cross section predictions for the same lens disc galaxy in the MOND scenario and the standard DM paradigm. The DM halo is assumed to be a conventional Non-Singular Isothermal Sphere \citep[NSIS, see e.g.,][]{Kormann_1994}
\begin{equation}
    \rho_{\text{NSIS}}(r) = \frac{\rho_0r_0^2}{\left(r^2+r_0^2\right)}\, ,
\end{equation}
with parameters taken to be $\rho_0 = 6.36\cdot 10^7\, \text{M}_\odot/\text{kpc}^3$, and $r_0 = 3.44$ kpc, for the template disc galaxy with $\Md = 10^{11}\, \text{M}_\odot$, $\Rd = 3.5\, \text{kpc}$, $\zd = 0.35\, \text{kpc}$, $\Mb = 10^{9}\, \text{M}_\odot$, and $\rb = 0.7\, \text{kpc}$. The DM halo parameters are obtained by fitting the Newtonian rotation curve of the dark matter-baryon system to the MOND-derived rotation curve. Namely, we are ``normalising'' the DM model to obtain a rotation curve matching the MOND prediction. Here, to extrapolate the MONDian rotation curve on the galactic plane, $v_{\text{ROT}}$, we directly employ the known algebraic MOND formula \citep{Banik_2018}, i.e., 
\begin{equation}
   v_{\text{ROT}}(R,0) = \sqrt{ R \, \left[1+\tilde{\nu} \left( y(R,0)\right)\right]\, |\partial_R\Phi_{\text{N}}(R,0)| } \, ,
\end{equation} 
Fig. \ref{fig:rot_curve} shows the MONDian rotation curve and the best fit for the chosen DM halo model, respectively. Fig.\ \ref{fig:MONDvDM_inc} shows the clear difference between the predictions of the two frameworks, with MOND SGL cross sections being distinctly higher than the DM alternative. Whilst in the DM paradigm inclination effects are present, these are increased in the MOND scenario, as we could expect from the presence of a ``phantom dark disc'' in the PDM distribution (see Fig. \ref{fig:rho_density}). Here we note that, a priori, a different choice of halo model, e.g., either the Navarro--Frenk--White profile \citep{Navarro_1996a, Navarro_1996b}, or a Burkert halo \citep{Persic_1990, Persic_1996, Salucci_2000, Salucci_2019} could produce a different SGL effect. However, we find that changing the DM halo profile does not affect the lensing cross section in a relevant manner if we maintain a ``normalisation'' of the DM halo w.r.t.\ the expected MONDian rotation curve. Therefore, we conclude that disc galaxy SGL is able to differentiate between the two paradigms. 

\section{Conclusions}\label{sec:Conclusions}
The dark matter framework has been highly successful in describing a plethora of phenomena, ranging from astrophysical to cosmological observations. Nonetheless, the nature of dark matter still eludes the community and currently represents one of the greatest mysteries in physics. As such, alternatives to the dark matter paradigm, which are able to match a wide-variety of observations, warrant investigation. Within the landscape of alternatives to dark matter, MOND is often considered empirically as the strongest contender to the dark matter scenario. Indeed, MOND has passed numerous observational tests and also produced a variety of now confirmed predictions.

Strong gravitational lensing by galaxies is capable of investigation in both regions of high, and low Newtonian accelerations. Therefore, in principle, it represents a valuable probe for MONDian effects over the full range of the acceleration scales within galaxies. In particular, studies of disc galaxies would enable a coupling of rotation curve dynamics and strong gravitational lensing. Hence, these represent a great laboratory for investigating gravitational physics. 

In this paper, we have derived the strong gravitational lensing effects of disc galaxies in MOND. In the QUMOND formulation of MOND we have employed the concept of equivalent Newtonian systems to apply the standard lensing formalism to MOND. Within this framework, we obtained the highly non-trivial distribution of phantom dark matter for disc galaxies and probed its characteristics. 

We have found that disc galaxy lensing in MOND produces physically plausible values of deflection angle, magnification and shear. Furthermore, we have shown that the inclination of the disc galaxy lens w.r.t.\ the observer highly influences the lensing geometry. We have seen that within the MOND paradigm, the classical caustic curves' morphology and number of lensed images expected for disc lenses is readily recovered. Moreover, we have probed the impact of lens morphological and mass parameters on the strong lensing cross section and found non-trivial, and even counter-intuitive, correlations due to the onset of purely MONDian effects.

Furthermore, we have drawn comparisons between inclination effects on the lensing cross section predicted in MOND and the dark matter framework. We have found that MOND non-linearities dominate over standard Newtonian lensing effects, resulting in: \textbf{(i)} steeper inclination effects (correlated to the non-spherical phantom dark matter density profile); and \textbf{(ii)} overall larger lensing cross sections.
In particular, it is noteworthy that a difference in the cross section of the obtained magnitude (which directly correlates to lensing likelihood) will still clearly impact the number of disc galaxy lenses predicted to be observed in upcoming optical surveys such as Euclid and LSST. Therefore, these future surveys will be able to distinguish the MOND and dark matter paradigms purely by disc galaxy gravitational lensing number counts. 

Our results might still need careful consideration, as they are ultimately dependent on the choice of interpolating function of MOND. In this paper, we have focused on the widely used form of interpolating function in Eqn.\ \eqref{eq:our_nu}, which has given a realisation of MOND able to pass a plethora of observational tests. We have also employed the alternative radial acceleration relation interpolating function (see Appendix B) to check the robustness of our results. We find that these interpolating functions produce similar lensing signatures, and both are distinct from the dark matter predictions. Therefore, although a different choice of interpolating function could ultimately produce different lensing effects, we believe our results will generalise to a broad range of choices of interpolating function. It is then exactly this (limited) dependence on the interpolating function that makes strong gravitational lensing by disc galaxies an exciting astrophysical tool to probe and constrain the full applicability of MOND.

\section*{Acknowledgements}
The authors acknowledge the partial support for this work by the Marsden Fund administered by the Royal Society of New Zealand, Te Apārangi under grant M1271. The authors are incredibly grateful to John Forbes for his valuable insights on the phenomenology of disc galaxies and clearing so many of our doubts. We deeply thank Zachary Lane and Michael Williams for their irreplaceable expertise in navigating computational issues, and Emma J. Johnson for producing the excellent plot in Fig.\ \ref{fig:lensing_geometry}. We are profoundly grateful to Pavel Kroupa and Indranil Banik for their insights on the MONDian landscape. We thank Morag Hills, Pierre Mourier, and David L. Wiltshire for their comments on a first draft of this paper. Finally, we are grateful to Chris Stevens, Ryan Ridden-Harper, and Shreyas Tiruvaskar for useful discussions. Finally, we are grateful to the anonymous reviewer for their comments and suggestions, which have helped improving this paper.  

\bibliography{references}{}
\bibliographystyle{aasjournal}

\appendix
\section{Code Routine}

The code routine developed to obtain the results in this work is available at the GitHub repository -- \href{https://github.com/chrisharhaw/MOND_lensing.git}{https://github.com/chrisharhaw/MOND\_lensing.git}. Here, we give a brief explanation of the logic implemented in the three main codes of the routine, i.e., \texttt{density\_grid.py}, \texttt{alpha\_rad\_tang.py}, and \texttt{caustic\_area.py}.

\subsection{\mbox{\texttt{density\_grid.py}}}

\begin{figure*}[htb]
    \centering
    \includegraphics[width=0.5\textwidth]{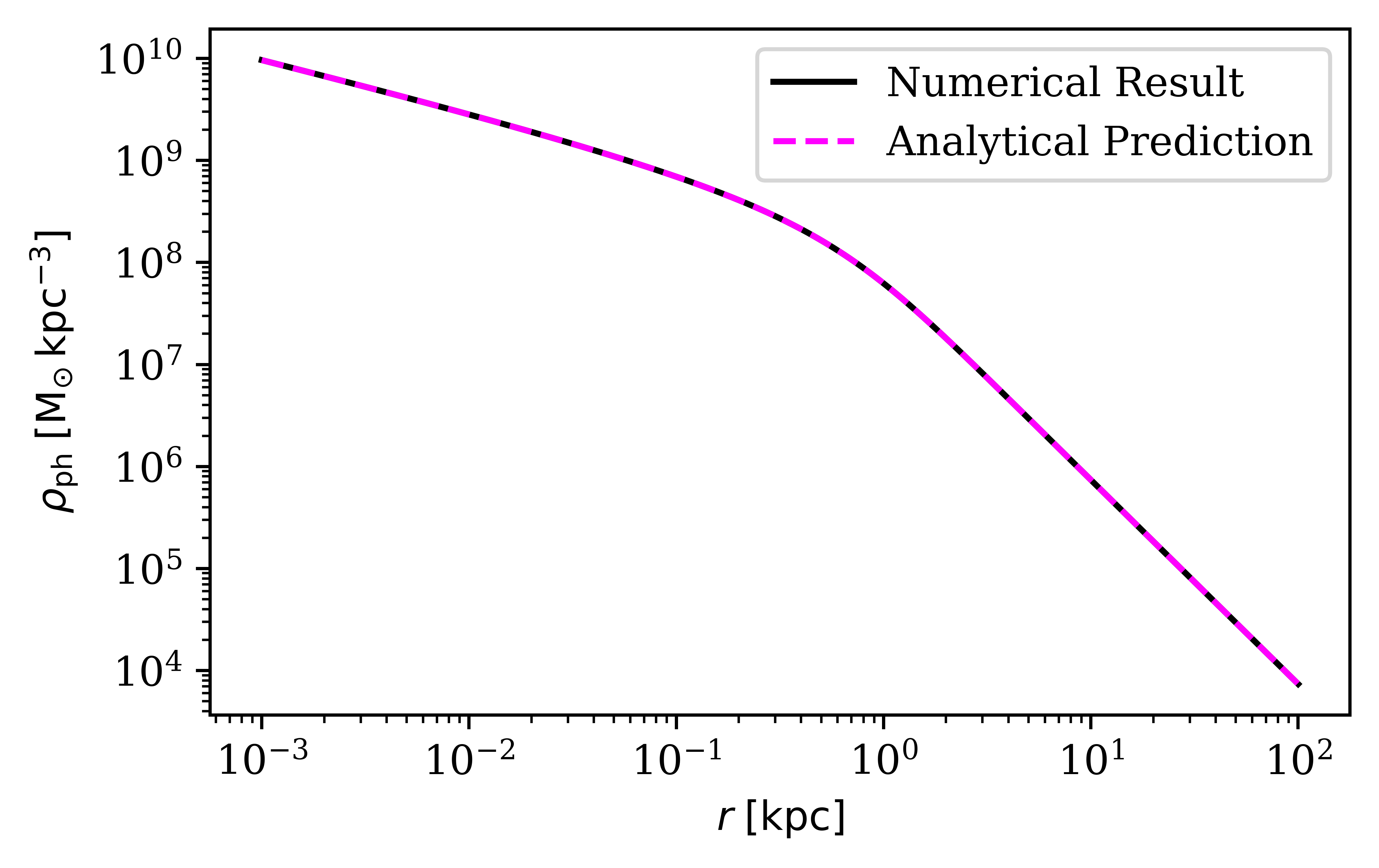}
    \caption{A comparison between the numerically obtained PDM (black line) and analytical prediction (dashed purple line) for a Plummer profile with $\text{M}_\odot$, and $\rb = 0.7\, \text{kpc}$. The MOND interpolating function selected is the one displayed in Eq.\ \eqref{eq:our_nu}.}
    \label{fig:SimpleCheck}
\end{figure*}

This code generates the baryonic and PDM distribution on a $(R,z)$ grid associated with a choice of disc galaxy parameters, and of grid boundaries and step. To calculate the PDM we implement the following steps 
\begin{enumerate}
    \item Define the Newtonian potentials for the exponential disc and the spherical Plummer bulge according to Eqs.\ \eqref{eq:PhiNd} and \eqref{eq:PhiNb}.
    \item Define the gradients in cylindircal coordinates of the potentials, namley
    \begin{align}
           &\vec{\nabla} \Phi_{\text{N},\text{d}} = G \Md \left( \left[ \int^\infty_0 \text{d}k \frac{J_1(R k) k}{(1 + \Rd^2 k^2)^{3/2}} Z(z,k) \right] \hat{R} + \hat{z} \left[ - \int^\infty_0 \text{d}k \frac{J_0(Rk)}{(1 + \Rd^2 k^2)^{3/2}} \frac{dZ}{dz} \right] \right) \, ,\\
   &\vec{\nabla} \Phi_{\text{N},\text{b}} = \frac{\Mb G}{(\rb^2 + R^2)^{3/2}} (R \hat{R} + z \hat{z }) \,,
    \end{align}
    where we have introduced the function
    \begin{equation}
        Z(z,k) := \frac{e^{-|z| k} - z_d k e^{-|z|/z_d}}{1 - z_d^2 k^2}\, .
    \end{equation}
    \item Compute the gradient of the total Newtonian potential by direct sum of $\vec{\nabla} \Phi_{\text{N},\text{d}}$, and $\vec{\nabla} \Phi_{\text{N},\text{b}}$, i.e., $\vec{\nabla} \Phi_{\text{N}} =\vec{\nabla} \Phi_{\text{N},\text{d}}+\vec{\nabla} \Phi_{\text{N},\text{b}}$.
    \item Employ the definition of PDM of Eq.\ \eqref{eq:PDM} in the following equivalent form
    \begin{equation}
        \rho_{\text{ph,b}} = \frac{1}{4\pi G} \frac{d\tilde{\nu}}{dy}\left(\vec{\nabla}y\cdot \vec{\nabla}\Phi_{\text{N}}\right) + \tilde{\nu}(y)\rho_{\text{B}}\, , 
    \end{equation}
    where we compute $\vec{\nabla}y$ directly in terms of the norm of $\vec{\nabla} \Phi_{\text{N}}$. Here, $\tilde{\nu}$ and its derivative are computed once a functional form is given.
\end{enumerate}
Finally, as output the code also returns the sum of the baryon and PDM densities, multiplied by a weight factor $R$ --accounting for part of the volume element contribution to the calculation of the deflection angle in \texttt{alpha\_rad\_tang.py}, and thus compensating for divergences in the PDM close to the symmetry axis. All of the integration of non-standard integrals are carried over using the \texttt{quad}, and \texttt{quad\_vec} functions of the python package \texttt{integrate} \citep{Virtanen_2020}. Fig. \ref{fig:SimpleCheck} shows the comparsion between the PDM density obtained with our code for a Plummer profile ($\text{M}_\odot$, and $\rb = 0.7\, \text{kpc}$), and the respective analytical prediction, for the choice of interpolating function of Eq.\ \eqref{eq:our_nu}. We find the two to be in good agreemnt.

\subsection{\mbox{\texttt{alpha\_rad\_tang.py}}}
This code loads the output of \texttt{density\_grid.py} and computes the deflection angles' components. To do so, we exploit an interesting formulation of the deflection angle. From Eqs.\ \eqref{eq:alpha}-\eqref{eq:Nprop} we can write
\begin{equation}
    \vec{\hat{\alpha}}(\xi_1,\xi_2) = \frac{2G}{c^2} \int^\infty_{-\infty} \text{d}\xi_3 \iiint \frac{ \rho(\vec{r})}{|\vec{\xi}-\vec{r}|} \, \text{d}^3 \vec{r} \,.
\end{equation}
Noting that the dependence on $\xi_3$ within the integral is present exclusively in the geometric term of the integrand (namely, the denominator), we swap the integration order to obtain
\begin{equation}
    \vec{\hat{\alpha}}(\xi_1,\xi_2) = \frac{2G}{c^2} \iiint \rho(\vec{r}) \text{d}^3 \vec{r}\int^\infty_{-\infty} \frac{ \text{d}\xi_3}{|\vec{\xi}-\vec{r}|} \, .
    \label{eq:mid_step}
\end{equation}
The second intergal in Eq. \eqref{eq:mid_step} is readily computed to obtain the two components of the deflection angle
\begin{align}
    &\alpha_{\xi_1} = \frac{2G}{c^2}\iiint \left(\frac{2\left(z\sin\lambda+x\cos\lambda-\xi_1\right)}{\xi^2_1+(y-\xi_2)^2+(z\sin\lambda+x\cos\lambda)^2-2\xi_1(z\sin\lambda+x\cos\lambda)}\right) \rho(\vec{r}) \text{d}^3 \vec{r} \, , \label{eq:alphaxi1}\\
    & \alpha_{\xi_1} = \frac{2G}{c^2}\iiint \left(\frac{y-\xi_2}{\xi^2_1+(y-\xi_2)^2+(z\sin\lambda+x\cos\lambda)^2-2\xi_1(z\sin\lambda+x\cos\lambda)}\right) \rho(\vec{r}) \text{d}^3 \vec{r} \, , \label{eq:alphaxi2}
\end{align}
where $\lambda$ defines the angle between $\xi_3$ and the $z$ axis of the coordinate system adapted to the disc galaxy mass distribution. Then, in this code we employ Eqs. \eqref{eq:alphaxi1} and \eqref{eq:alphaxi2} to define the components of the deflection angle. We compute these integrals by direct discretisation on a grid, whose precision can be specified. From the deflection angle components, the code also computes the inverse magnification matrix and extracts its radial and tangential eigenvalues.

\subsection{\mbox{\texttt{caustic\_area.py}}}
This code loads the outputs of \texttt{alpha\_rad\_tang.py}, i.e., the deflection angle components, and the tangential and radial eigenvalues of the inverse magnification matrix, $\lambda_{^t_r}$. The radial and tangential critical curves can then be computed by determining the corresponding contours where $\lambda_{^t_r} = 0$, which are traced back into caustic lines using the lens equation (Eq.~\eqref{eq:lens_eq}). Once the tangential caustic contours have been found (if they exist), the area within them is calculated. We do this for each density and inclination being investigated, then finally output these values as a \texttt{numpy} array for plotting.

\section{The RAR interpolating function}
\begin{figure*}[htbp]
    \begin{minipage}[b]{0.5\textwidth}
        \includegraphics[width=\linewidth]{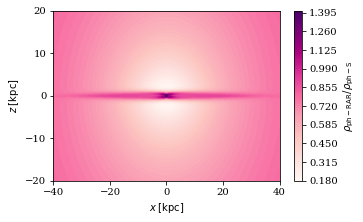}
    \end{minipage}
    \begin{minipage}[b]{0.5\textwidth}
        \includegraphics[width=\linewidth]{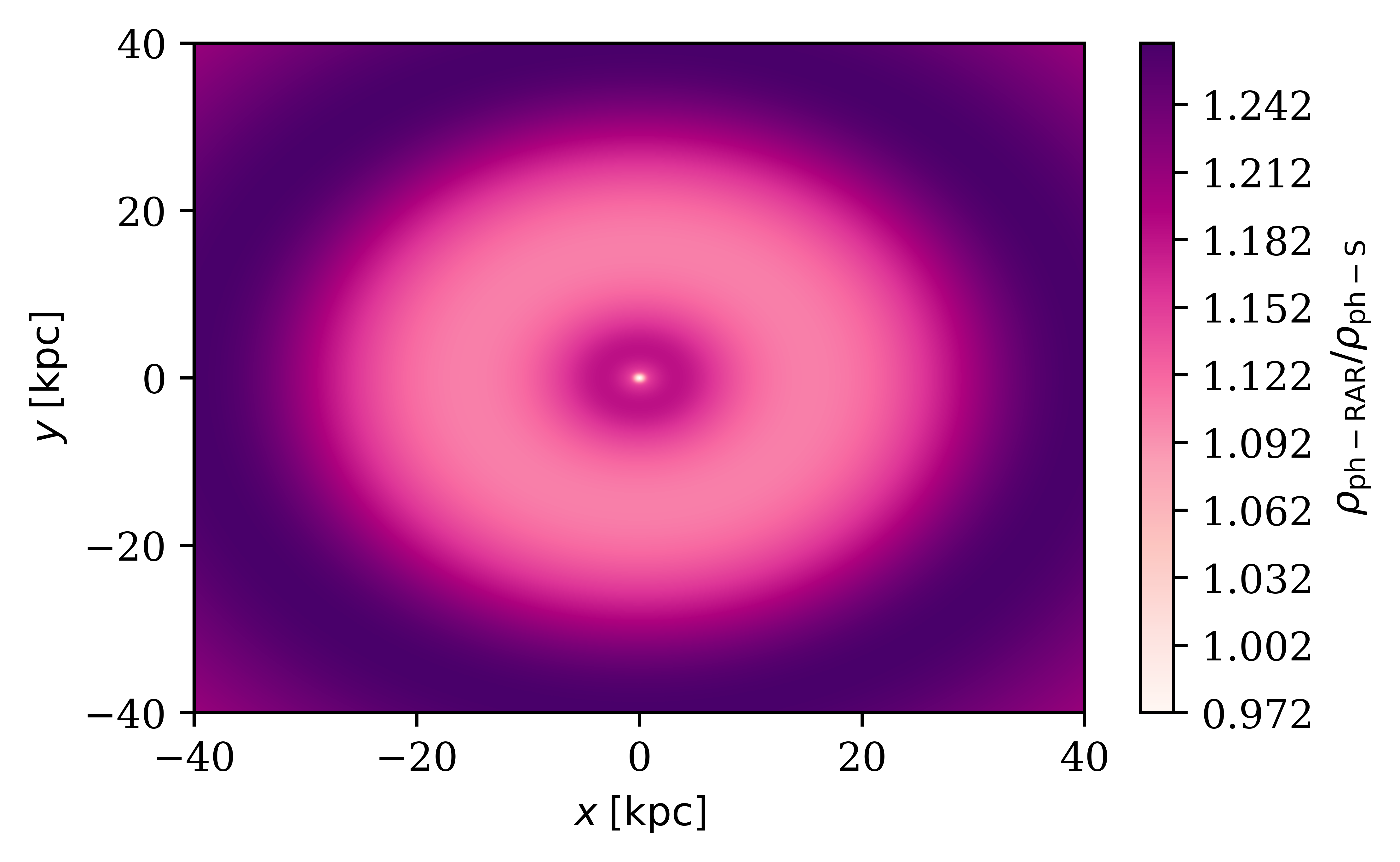}
    \end{minipage}
    \caption{PDM densities ratio, $\rho_{\text{ph-RAR}}/\rho_{\text{ph-S}}$ on the $z-x$ plane (left panel) and on the $x-y$ plane (right panel), respectively, for a disc galaxy with $\Md = 10^{11}\, \text{M}_\odot$, $\Rd = 3.5\, \text{kpc}$, $\zd = 0.35\, \text{kpc}$, $\Mb = 10^{9}\, \text{M}_\odot$, and $\rb = 0.7\, \text{kpc}$.}
    \label{fig:ratio_density_RAR}
\end{figure*}

To check the robustness of our result w.r.t.\ the change of the interpolating function, we consider a comparison with the RAR interpolating function \citep{McGaugh_2016,Stiskalek_2023}, i.e., 
\begin{equation}
    \tilde{\nu}_{\text{RAR}}(y) := \frac{e^{-\sqrt{y}}}{1-e^{-\sqrt{y}}} \, .
\end{equation}
We computed the RAR PDM distribution for the template disc galaxy model -- namely, $\Md = 10^{11}\, \text{M}_\odot$, $\Rd = 3.5\, \text{kpc}$, $\zd = 0.35\, \text{kpc}$, $\Mb = 10^{9}\, \text{M}_\odot$, and $\rb = 0.7\, \text{kpc}$ -- and find that the obtained PDM distribution is overall compatible with the one derived with the interpolating function of Eq.~\eqref{eq:our_nu}, henceforth referred to as $\rho_{\text{ph-RAR}}$, and $\rho_{\text{ph-S}}$, respectively. In Fig.\ \ref{fig:ratio_density_RAR} we show the face-on view, and side view of the PDM densities ratios, i.e., $\rho_{\text{ph-RAR}}/\rho_{\text{ph-S}}$, for the chosen template galaxy. We see that although the two PDM densities display differences, these give corrections of one w.r.t.\ the other only of $\mathcal{O}(1)$. Here, we note that in front of $\mathcal{O}(1)$ differences between the two distributions, the ratios between the two PDM densities display non-trivial morphological features, charting the different behaviour of two interpolating functions. 

\begin{figure*}
    \centering
    \includegraphics[width=0.8\textwidth]{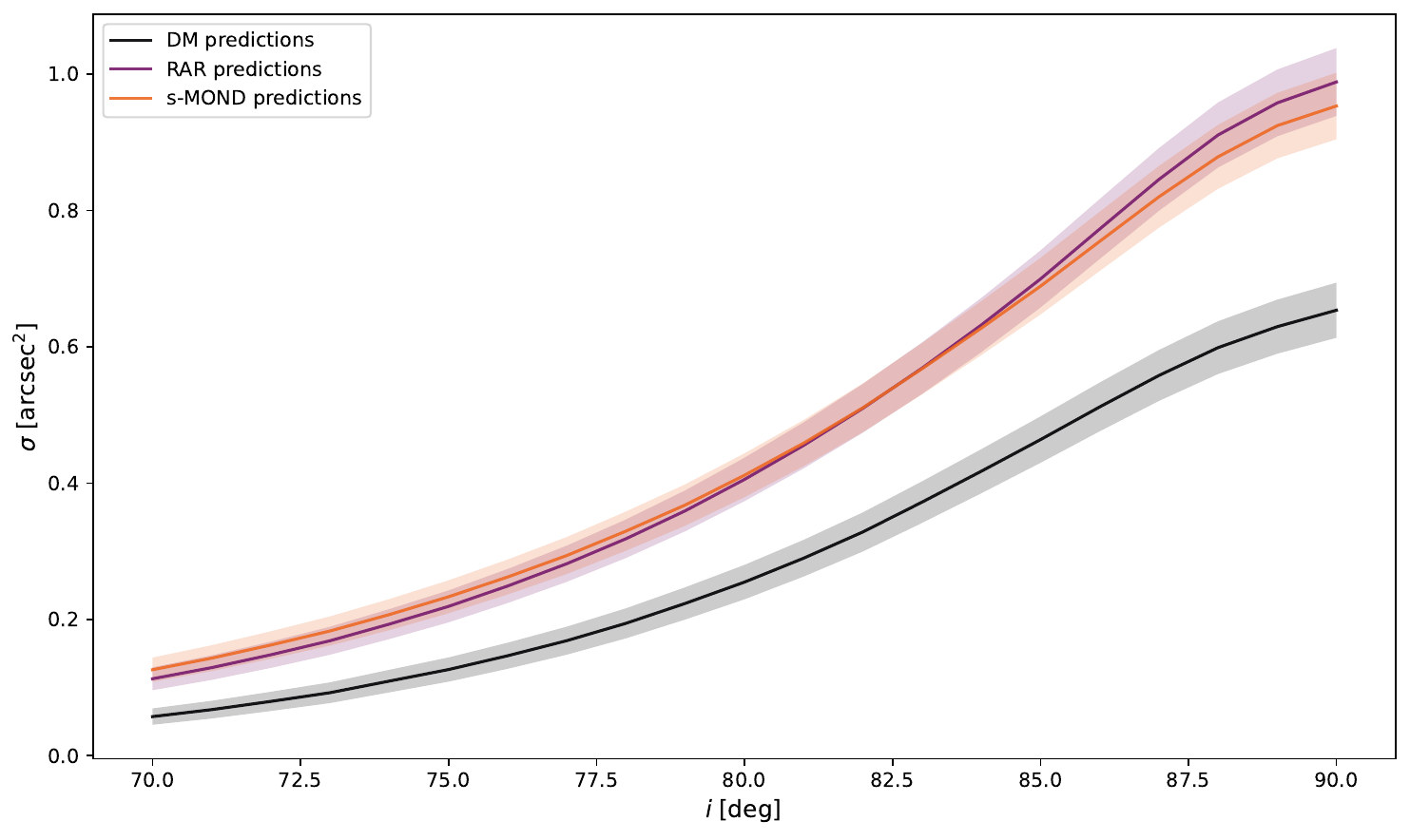}
    \caption{A comparison of the effects of varying inclination on the SGL cross section produced for the interpolating function used throughout this work, the RAR function, and the DM halo alternative. The disc galaxy parameters are $\Md = 10^{11}\, \text{M}_\odot$, $\Rd = 3.5\, \text{kpc}$,  $\zd = 0.35\, \text{kpc}$, and $\Mb = 10^{9}\, \text{M}_\odot$, and $\rb = 0.7\, \text{kpc}$. The DM halo is taken as a NSIS profile with parameters $\rho_0 = 6.36\cdot 10^7\, \text{M}_\odot/\text{kpc}^3$, and $r_0 = 3.44$ kpc. The lens and source are placed at redshift $z_{\text{L}} = 0.5$ and $z_{\text{S}} = 2.0$, respectively. The shaded areas represent the one standard deviation uncertainties around the obtained SGL cross section curves.}
    \label{fig:RAR_caustic_inc}
\end{figure*}

Finally, we have computed the SGL expected for the template galaxy whilst employing the RAR interpolation function. Fig.\ \ref{fig:RAR_caustic_inc} shows a comparison between the inclination effects on the SGL cross section produced for the interpolating function used throughout this work, the RAR function, and the DM halo alternative. As we could expect from the obtained ratios between the two PDM distributions, the SGL effects obtained for the two choices of interpolating functions are almost identical, and distinguishable from the DM alternative.


\end{document}